\documentclass[aps,prl,twocolumn,preprintnumbers,nofootinbib,amsmath,amssymb]{revtex4-2}
\usepackage{graphicx}
\usepackage{dcolumn}
\usepackage{bm}
\usepackage{amsmath}
\usepackage{braket}
\usepackage[hidelinks]{hyperref}
\usepackage{epstopdf}
\usepackage{placeins}
\usepackage{multirow}
\usepackage{makecell}
\usepackage{cleveref}
\usepackage{xcolor}
\usepackage{dblfloatfix}

\newcommand{\beq}{\begin{eqnarray}}
\newcommand{\eeq}{\end{eqnarray}}
\newcommand{\beqnn}{\begin{eqnarray*}}
\newcommand{\eeqnn}{\end{eqnarray*}}
\newcommand{\Tr}{\ensuremath{\mathrm{Tr}}}

\newcommand{\cool}{\mathrm{cool}}
\newcommand{\QCD}{\mathrm{QCD}}
\newcommand{\sphal}{\mathrm{Sphal}}
\newcommand{\Cov}{\mathrm{Cov}}

\renewcommand\thesection{\arabic{section}}

\usepackage[font=small, labelfont=bf, textfont={small,it}]{caption}
\def\spose#1{\hbox to 0pt{#1\hss}}
\def\ltapprox{\mathrel{\spose{\lower 3pt\hbox{$\mathchar"218$}}
\raise 2.0pt\hbox{$\mathchar"13C$}}}

\graphicspath{{figs/}}

\begin{document}

\title{Sphaleron rate of $N_f=2+1$ QCD}

\author{Claudio Bonanno}
\email{claudio.bonanno@csic.es}
\affiliation{Instituto de F\'isica Te\'orica UAM-CSIC, c/ Nicol\'as Cabrera 13-15, Universidad Aut\'onoma de Madrid, Cantoblanco, E-28049 Madrid, Spain}

\author{Francesco D'Angelo}
\email{francesco.dangelo@phd.unipi.it}
\affiliation{Dipartimento di Fisica dell'Universit\`a di Pisa \& \\ INFN Sezione di Pisa, Largo Pontecorvo 3, I-56127 Pisa, Italy}

\author{Massimo D'Elia}
\email{massimo.delia@unipi.it}
\affiliation{Dipartimento di Fisica dell'Universit\`a di Pisa \& \\ INFN Sezione di Pisa, Largo Pontecorvo 3, I-56127 Pisa, Italy}

\author{Lorenzo Maio}
\email{lorenzo.maio@phd.unipi.it\\}
\affiliation{Dipartimento di Fisica dell'Universit\`a di Pisa \& \\ INFN Sezione di Pisa, Largo Pontecorvo 3, I-56127 Pisa, Italy}
\affiliation{Aix Marseille Univ., Université de Toulon, CNRS, CPT, Marseille, France.}

\author{Manuel Naviglio}
\email{manuel.naviglio@phd.unipi.it}
\affiliation{Dipartimento di Fisica dell'Universit\`a di Pisa \& \\ INFN Sezione di Pisa, Largo Pontecorvo 3, I-56127 Pisa, Italy}

\date{\today}

\begin{abstract}
We compute the sphaleron rate of $N_f=2+1$ QCD at the physical point for a range of temperatures $200$~MeV~$\lesssim T \lesssim 600$~MeV. We adopt a strategy recently applied in the quenched case, based on the extraction of the rate via a modified version of the Backus-Gilbert method from finite-lattice-spacing and finite-smoothing-radius Euclidean topological charge density correlators. The physical sphaleron rate is finally computed by performing a continuum limit at fixed physical smoothing radius, followed by a zero-smoothing extrapolation. Dynamical fermions were discretized using the staggered formulation, which is known to yield large lattice artifacts for the topological susceptibility. However, we find them to be rather mild for the sphaleron rate.
\end{abstract}

\maketitle

\section{Introduction}

The rate of real-time QCD topological transitions, the so-called strong sphaleron rate,
\beq\label{eq:rate_def}
\begin{aligned}
\Gamma_\sphal &= \underset{t_{\mathrm{M}}\to\infty}{\underset{V_s\to\infty}{\lim}} \, \frac{1}{V_s t_{\mathrm{M}}}\left\langle\left[\int_0^{t_{\mathrm{M}}} d t_{\mathrm{M}}' \int_{V_s} d^3x \, q(t_{\mathrm{M}}', \vec{x})\right]^2\right\rangle\\
\\[-1em]
\\[-1em]
\\[-1em]
&=\int d t_{\mathrm{M}} d^3x \braket{q(t_{\mathrm{M}},\vec{x}) q(0,\vec{0})},
\end{aligned}
\eeq
where $t_\mathrm{M}$ is the Minkowski time and $q = (\alpha_s/8\pi)G\widetilde{G}$ is the QCD topological charge density, plays a crucial role in several phenomenological contexts.

For example, during heavy-ion collisions, where a hot medium of quarks and gluons and strong magnetic fields are created for a short time, a non-vanishing sphaleron rate in the quark-gluon plasma can create local imbalances in the number of left and right-handed quark species, leading in particular to the so-called Chiral Magnetic Effect~\cite{Fukushima:2008xe, Kharzeev:2013ffa, Laine:2011xm, Astrakhantsev:2019zkr, Almirante:2023wmt}, i.e., the appearing of an electric current flowing in the quark-gluon medium in the parallel direction to the magnetic field.

Another example is offered by axion phenomenology, where recently it has been argued that the strong sphaleron rate plays an intriguing role~\cite{Notari:2022ffe}. As a matter of fact, the QCD strong sphaleron rate describes the rate of axion creation/annihilation in the early Universe, and such quantity directly enters the Boltzmann equation for the time-evolution of the axion number distribution in the cosmological medium.

It is therefore clear that a first-principle and fully non-perturbative computation of the QCD sphaleron rate at finite temperature constitutes an essential input to provide fundamental phenomenological predictions about the Standard Model and beyond. However, so far results in the literature have been limited to the quenched case, i.e., the pure-gauge Yang--Mills theory~\cite{Kotov:2018aaa, Kotov:2019bt, Altenkort:2020axj, BarrosoMancha:2022mbj,Bonanno:2023ljc}.

In this letter we present a first non-perturbative determination of the sphaleron rate in $2+1$ QCD at the physical point from numerical Monte Carlo simulations on the lattice above the chiral crossover. In particular, we explored a temperature range $200$~MeV~$\lesssim T \lesssim 600$~MeV, and the rate was computed adopting the strategy we recently applied in the quenched case in~\cite{Bonanno:2023ljc}.

\section{Methods}\label{sec:methods}
We performed Monte Carlo simulations of $N_f=2+1$ QCD at the physical point for five temperatures: $T=230, 300, 365, 430$ and $570$ MeV. For each temperature, we explored 3--5 values of the lattice spacing, keeping the physical lattice volume constant and choosing the bare coupling and the bare quark masses so as to move on a Line of Constant Physics (LCP), where $m_s/m_l = 28.15$ and $m_\pi\simeq135$~MeV were kept constant and equal to their physical value~\cite{Aoki:2009sc,Borsanyi:2010cj,Borsanyi:2013bia}. The gauge sector has been discretized by using the tree-level Symanzik improved Wilson gauge action, while the quark sector was discretized adopting rooted stout staggered fermions.

Gauge configurations have been generated adopting the standard Rational Hybrid Monte Carlo (RHMC) updating algorithm, used in combination with the multicanonical algorithm. Above the QCD chiral crossover $T_c\simeq155$~MeV, the topological susceptibility $\chi\equiv\frac{\braket{Q^2}}{V}$, $Q=\int d^4x~q(x)$, is suppressed as a power-law of the temperature~\cite{Petreczky:2016vrs,Borsanyi:2016ksw,Lombardo:2020bvn,Athenodorou:2022aay}. Due to such suppression, on typical lattice volumes $\braket{Q^2} = V \chi \ll1$, thus topological fluctuations are suppressed and the probability distribution of $Q$ is dominated by $Q=0$. Thus, large statistics are needed to properly sample the topological charge distribution. The multicanonical algorithm allows to easily bypass this issue by adding a topological bias potential to the gauge action that enhances the probability of visiting suppressed topological sectors, without spoiling importance sampling. Expectation values with respect to the original path-integral probability distribution are then exactly recovered via a standard reweighting~\cite{Jahn:2018dke,Bonati:2018blm,Athenodorou:2022aay,Bonanno:2022dru}.

The first step to determine the sphaleron rate is to obtain Euclidean lattice topological charge density correlators. The charge density was discretized using the standard gluonic clover definition:
\beq\label{eq:topcharge_dens_clover}
q_L(n) = \frac{-1}{2^9 \pi^2}\sum_{\mu\nu\rho\sigma=\pm1}^{\pm4}\varepsilon_{\mu\nu\rho\sigma}
\Tr\left\{\Pi_{\mu\nu}(n)\Pi_{\rho\sigma}(n)\right\},
\eeq
where $\Pi_{\mu\nu}(n)$ is the plaquette and $\varepsilon_{(-\mu)\nu\rho\sigma} = - \varepsilon_{\mu\nu\rho\sigma}$.

We first compute the time profile $Q_L(n_t)$ of the topological charge $Q_L$:
\beq\label{eq:topcharge_profile}
Q_L(n_t) = \sum_{\vec{n}} q_L(n_t,\vec{n}), \quad Q_L = \sum_n q_L(n).
\eeq
Then, we obtain the topological charge density correlator in dimensionless physical units as:
\beq\label{eq:topchargedens_lat}
\frac{G_L(tT)}{T^5} = \frac{N_t^5}{N_s^3}\braket{Q_L(n_{t,1})Q_L(n_{t,2})},
\eeq
where $N_s$ and $N_t$ are the spatial and temporal extents of the lattice and
\beq
tT=\min\left\{\vert n_{t,1} - n_{t,2} \vert/ N_t;~1-\vert n_{t,1} - n_{t,2} \vert/ N_t\right\}.
\eeq
is the physical time separation between the sources entering the correlator.

The topological charge profiles are computed on smoothened configurations. Smoothing is used to dampen UV fluctuations affecting the two-point function of the correlator of the lattice topological charge density, which would otherwise result in additive and multiplicative renormalizations~\cite{DiVecchia:1981aev,Campostrini:1988cy,DElia:2003zne,Vicari:2008jw}. Several smoothing algorithms have been proposed, e.g., cooling~\cite{Berg:1981nw,Iwasaki:1983bv,Itoh:1984pr,Teper:1985rb,Ilgenfritz:1985dz,Campostrini:1989dh,Alles:2000sc}, stout smearing~\cite{APE:1987ehd, Morningstar:2003gk} or gradient flow~\cite{Luscher:2009eq, Luscher:2010iy}, all agreeing when properly matched to each other~\cite{Alles:2000sc, Bonati:2014tqa, Alexandrou:2015yba}. In this work we adopt cooling for its numerical cheapness. A single cooling step consists in aligning each link to its relative staple, so that the local action density is minimized.

Concerning the rate computation, we recall that Eq.~\eqref{eq:rate_def} is of no use on the lattice, being it expressed in terms of Minkowskian correlators. However, the Kubo equation relates the sphaleron rate to the spectral density $\rho(\omega)$ of the Euclidean topological charge density correlator $G(t) = \int d^3 x \braket{q(x)q(0)}$ (here $t$ is the imaginary time)~\cite{Meyer:2011gj}:
\beq\label{eq:kubo}
\Gamma_\sphal = 2T \lim_{\omega \to 0} \frac{\rho(\omega)}{\omega},
\eeq
\beq\label{eq:rho_def}
G(t) = - \int_0^{\infty} \frac{d\omega}{\pi} \rho(\omega) \frac{\cosh\left[\omega(t-1/(2T))\right]}{\sinh\left[\omega/(2T)\right]}.
\eeq

Therefore, determining the sphaleron rate on the lattice translates into the problem of inverting the integral relation~\eqref{eq:rho_def} to compute $\rho(\omega)$ from lattice correlators $G_L(t)$. Strategies to solve inverse problems have been widely studied in the literature~\cite{Boito:2022njs,Horak:2021syv,DelDebbio:2021whr,Candido:2023nnb,Altenkort:2020axj,Altenkort:2020fgs,Altenkort:2022yhb,Altenkort:2023oms,Brandt:2015aqk,Brandt:2015sxa,Hansen:2019idp,Bulava:2021fre,ExtendedTwistedMassCollaborationETMC:2022sta,Frezzotti:2023nun,Evangelista:2023vtl,10.1093/gji/ggz520,Rothkopf:2022fyo,Aarts:2023vsf}. Here we rely on the recently-proposed modification~\cite{Hansen:2019idp} of the Backus--Gilbert inversion method~\cite{BackusGilbert1968:aaa}.

On general grounds, the Backus--Gilbert method assumes that the spectral density can be approximated via:
\beq\label{eq:Estimator}
\bar{\rho}(\bar{\omega}) = - \pi \bar{\omega} \sum_{t=0}^{1/T} g_t(\bar{\omega}) G(t),
\eeq
where the $g_t(\bar{\omega})$ are unknown coefficients that need to be determined. In our case, we are just interested in $\bar{\omega}=0$:
\beq
\left[\frac{\bar{\rho}(\bar{\omega})}{\bar{\omega}}\right]_{\bar{\omega}\,=\,0}= - \pi \sum_{t=0}^{1/T} g_t(0) G(t) = \frac{\Gamma_\sphal}{2T}.
\eeq
The determination of the $g_t$ coefficients is achieved through the minimization of a suitable functional. In particular we followed the strategy described in Ref.~\cite{Hansen:2019idp}, which was also the one we employed in the quenched case in Ref.~\cite{Bonanno:2023ljc}. Given the technicalities involved in such process, more details on this point are given in the Supplementary Material.

The last points to discuss are how to treat finite lattice spacing effects, and what is the impact of smoothing on the sphaleron rate. Let us start recalling that, after $n_\cool$ cooling steps are performed on the gauge fields, UV fluctuations are damped out, up to a distance known as the smoothing radius $r_s \propto a \sqrt{n_\cool}$. The first step, thus, is to take the continuum limit at fixed smoothing radius. In our setup, since $n_\cool \propto (r_s/a)^2$ and $N_t^{-1} = aT$, this means to keep $n_\cool/N_t^2 \propto (r_sT)^2$ constant for each lattice spacing. Assuming $O(a^2)$ corrections, we perform continuum extrapolations according to the fit function:
\beq
\Gamma_{\sphal,L}\left(N_t,\frac{n_\cool}{N_t^2}\right) = \Gamma_\sphal\left(\frac{n_\cool}{N_t^2}\right) + \frac{c}{N_t^2},
\eeq
where $\Gamma_{\sphal,L}(N_t, n_\cool)$ stands for the sphaleron rate obtained from the lattice correlator $G_L(tT)$ computed from a $N_s^3 \times N_t$ lattice after $n_\cool$ cooling steps.

In principle, one would expect a residual dependence of the continuum-extrapolated sphaleron rate on the smoothing radius, i.e., a residual dependence of $\Gamma_\sphal$ on $n_\cool/N_t^2$. However, for all temperatures we found that $\Gamma_\sphal$ is practically independent of $n_\cool/N_t^2$ for sufficiently small values of $n_\cool$. The same behavior was observed in the quenched theory~\cite{Bonanno:2023ljc}. Such evidence can be physically explained on the basis of the definition of the rate itself: the dominant contribution to $\rho(\omega)$ in the origin is given by the behavior of the topological charge density correlator at large time separations, and it is reasonable to expect it to be largely unaffected by the UV cut-off introduced by cooling, which mostly affects the short-distance behavior of $G_L(t)$.

\section{Results}\label{sec:rate_results}

\begin{table}[!t]
\begin{center}
\begin{tabular}{ |c|c|c|c|c|c|c|}
\hline
$T$~[MeV] & $T/T_c$ & $\beta$ & $a$~[fm] & $a m_s \cdot 10^{-2}$ & $N_s$ & $N_t$ \\
\hline
\multirow{5}{*}{230} & \multirow{5}{*}{1.48} & 3.814* & 0.1073 & 4.27 & 32 & 8  \\
&& 3.918* & 0.0857 & 3.43 & 40 & 10 \\
&& 4.014  & 0.0715 & 2.83 & 48 & 12 \\
&& 4.100  & 0.0613 & 2.40 & 56 & 14 \\
&& 4.181  & 0.0536 & 2.10 & 64 & 16 \\
\hline	
\multirow{4}{*}{300} & \multirow{4}{*}{1.94} & 3.938 & 0.0824 & 3.30 & 32 & 8  \\
&& 4.059  & 0.0659 & 2.60 & 40 & 10 \\
&& 4.165  & 0.0549 & 2.15 & 48 & 12 \\
&& 4.263  & 0.0470 & 1.86 & 56 & 14 \\
\hline	
\multirow{4}{*}{365} & \multirow{4}{*}{2.35} & 4.045 & 0.0676 & 2.66 & 32 & 8 \\
&& 4.175  & 0.0541 & 2.12 & 40 & 10 \\
&& 4.288  & 0.0451 & 1.78 & 48 & 12 \\
&& 4.377  & 0.0386 & 1.55 & 56 & 14 \\
\hline
\multirow{4}{*}{430} & \multirow{4}{*}{2.77} & 4.280 & 0.0458 & 1.81 & 32 & 10 \\
&& 4.385 & 0.0381 & 1.53 & 36 & 12 \\
&& 4.496 & 0.0327 & 1.29 & 48 & 14 \\
&& 4.592 & 0.0286 & 1.09 & 48 & 16 \\
\hline
\multirow{3}{*}{570} & \multirow{3}{*}{3.68} & 4.316  & 0.0429 & 1.71 & 32 & 8 \\
&& 4.459  & 0.0343 & 1.37 & 40 & 10 \\
&& 4.592  & 0.0286 & 1.09 & 48 & 12 \\
\hline
\end{tabular}
\end{center}
\caption{Summary of simulation parameters. The bare parameters $\beta$,
$a m_s$ and the lattice spacings $a$ have been fixed according to results of Refs.~\cite{Aoki:2009sc, Borsanyi:2010cj, Borsanyi:2013bia}, and the bare light quark mass $a m_l$ is fixed through $m_s/m_l=28.15$. Simulations marked with * have been performed without multicanonical algorithm as $\braket{Q^2}$ was sufficiently large to observe a reasonable number of fluctuations of the topological charge.}
\label{tab:simul_params}
\end{table}

\begin{table}[!t]
\begin{center}
\begin{tabular}{ |c|c|}
\hline
&\\[-1em]
$T$~[MeV] & $\Gamma_\sphal/T^4$ \\
&\\[-1em]
\hline
230 & 0.310(80)\\
300 & 0.165(40)\\
365 & 0.115(30)\\
430 & 0.065(20)\\
570 & 0.045(12)\\
\hline
\end{tabular}
\end{center}
\caption{Summary of the determinations of the sphaleron rate of $2+1$ full QCD at the physical point.}
\label{tab:rate_res}
\end{table}

All simulation points are summarized in Tab.~\ref{tab:simul_params}, while in Tab.~\ref{tab:rate_res} we summarize our results for the sphaleron rate as a function of the temperature.

\begin{figure}[!t]
\centering
\includegraphics[scale=0.47]{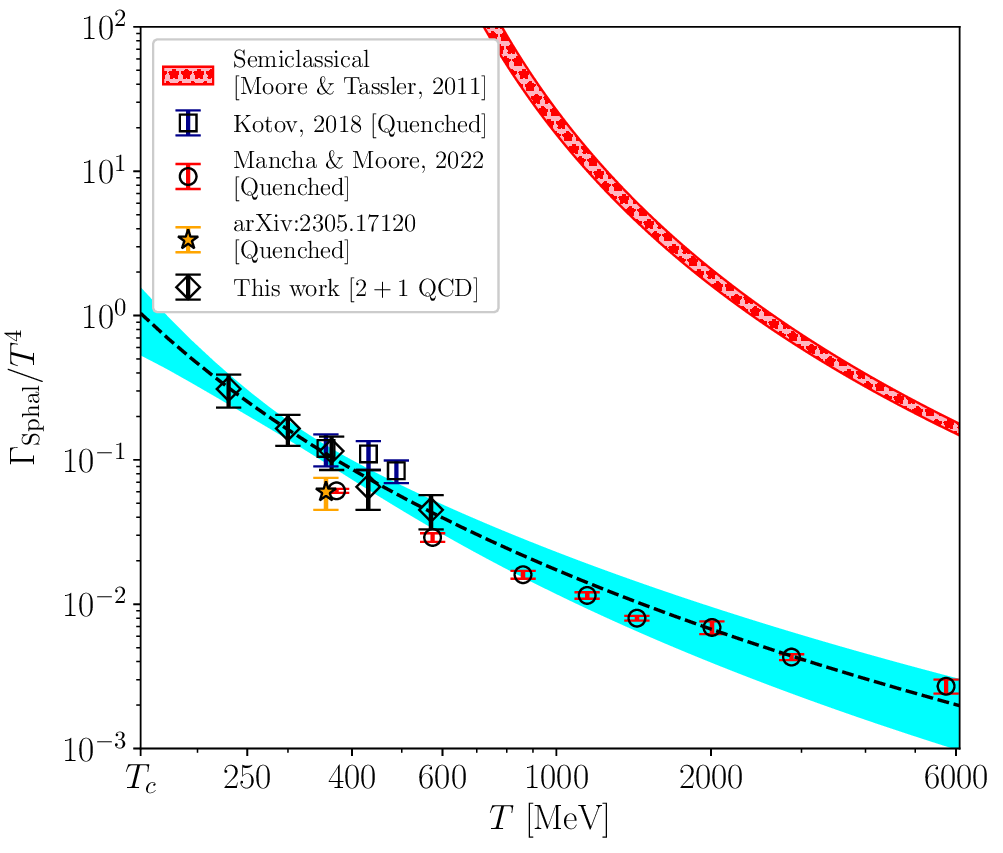}
\includegraphics[scale=0.47]{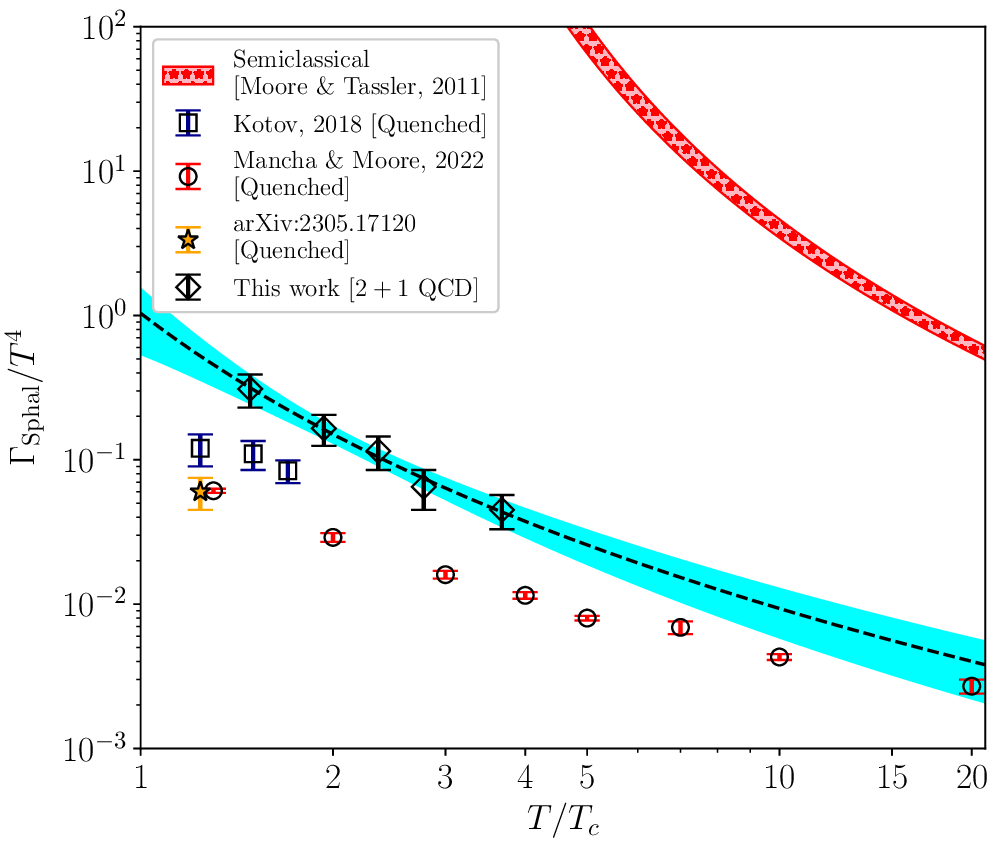}
\caption{Sphaleron rate for $2+1$ full QCD at the physical point as a function of temperature $T$ (diamond points). Dashed line and uniform shaded area represent best fit of our results according to~\eqref{eq:fit_sphal_ans1}. Previous quenched determinations of the rate are also shown: Refs.~\cite{Kotov:2018aaa,Kotov:2019bt} (square points), Ref.~\cite{BarrosoMancha:2022mbj} (round points) and Ref.~\cite{Bonanno:2023ljc} (starred point). Top plot: x-axis expressed in terms of absolute temperature $T$ converted in $\mathrm{MeV}$. Bottom plot: x-axis expressed in terms of $T/T_c$, where $T_c = 155~\mathrm{MeV}$ and $T_c=287$~MeV for full QCD and quenched results respectively. Starred shaded area depicts semiclassical prediction~\eqref{eq:semiclassic_result_quenched}.}
\label{fig:sphaleron_rate_fullQCD_ans1}
\end{figure}

First of all, we want to compare our full QCD results with the previous quenched determinations of~\cite{Kotov:2018aaa,BarrosoMancha:2022mbj,Bonanno:2023ljc}. Such comparison is shown in Fig.~\ref{fig:sphaleron_rate_fullQCD_ans1}. We observe that full QCD determinations turn out to be slightly larger (although of the same order of magnitude) than the quenched ones, both when we report the rates in terms of the absolute $T$ in MeV, and when we report them in terms of $T/T_c$, where for full QCD results we used the chiral crossover temperature $T_c=155$ MeV and for quenched results we used the critical temperature $T_c=287$ MeV.


We now move to the comparison of our results with available analytical predictions in the literature. In Refs.~\cite{Moore:2010jd,Berghaus:2020ekh}, the following semiclassical estimate for the sphaleron rate is reported:
\beq
\frac{\Gamma_\sphal}{T^4} \simeq C_1 \alpha_s^5,
\eeq
with $\alpha_s$ the running strong coupling.

Using the $1$-loop result for the temperature running of $\alpha_s(T)$ reported in~\cite{Steffens:2004sg}, one obtains:
\beq
\frac{\Gamma_\sphal}{T^4} &=& C_1 \left[\frac{C_2}{\log(T^2/\Lambda_\QCD^2)}\right]^5\\
\label{eq:semiclassic_result_quenched}
&\equiv& \left[\frac{A_0}{\log(T^2/T_c^2)+\log(B_0^2)}\right]^5,
\eeq
where $B_0 = T_c/\Lambda_\QCD \simeq 0.46(2)$ using the latest world-average FLAG value for the 3-flavor dynamically-generated scale $\Lambda_\QCD^{(\overline{\mathrm{MS}})}(\mu = 2~\mathrm{GeV}) \simeq 338(12)$~MeV~\cite{FlavourLatticeAveragingGroupFLAG:2021npn}, and where the overall pre-factor can be estimated to be $A_0 =C_1^{1/5}C_2 \simeq 3.08(2)$ using the expressions for $C_1$ and $C_2$ reported, respectively, in Refs.~\cite{Moore:2010jd,Steffens:2004sg}.


Inspired by the functional form of~\eqref{eq:semiclassic_result_quenched}, we performed a best fit of our data for $\Gamma_\sphal/T^4$ as a function of $T$ using the following fit function:
\beq\label{eq:fit_sphal_ans1}
\frac{\Gamma_\sphal}{T^4} = \left[\frac{A}{\log(T^2/T^2_c) + \log(B^2)}\right]^C.
\eeq
Our data are well described by Eq.~\eqref{eq:fit_sphal_ans1} with $C = 5$, with a reduced chi-squared of $0.36/3$. Actually, one should be cautious about this apparent success, in particular regarding the value of $C$. Indeed, the fit returns a similarly good value of the reduced chi-squared for a very large range of values of $C$, while if this parameter is left free, the best fit  returns a value $C = 5.6$ with a 100\% error. This is understandable, since our temperature range is still too small, and our statistical errors still too large, to get a precise estimate of the power of a logarithmic function.

If we fix $C$ to the value of the semiclassical prediction, we obtain the best fit depicted in Fig.~\ref{fig:sphaleron_rate_fullQCD_ans1} as a dashed line, 
while the uniform shaded area represents the corresponding error band; the fit parameters turn out to be $A=2.96(51)$ and $B=4.3(1.7)$. We note that we find a remarkable agreement for the pre-factor $A$ with the prediction $A_0$, while we find the pole parameter $B$ to be larger by an order of magnitude compared to $B_0$. This clarifies why the semiclassical prediction overestimates by more than 2 orders of magnitude the lattice data, cf.~Fig.~\ref{fig:sphaleron_rate_fullQCD_ans1}.

\begin{figure}[!t]
\centering
\includegraphics[scale=0.47]{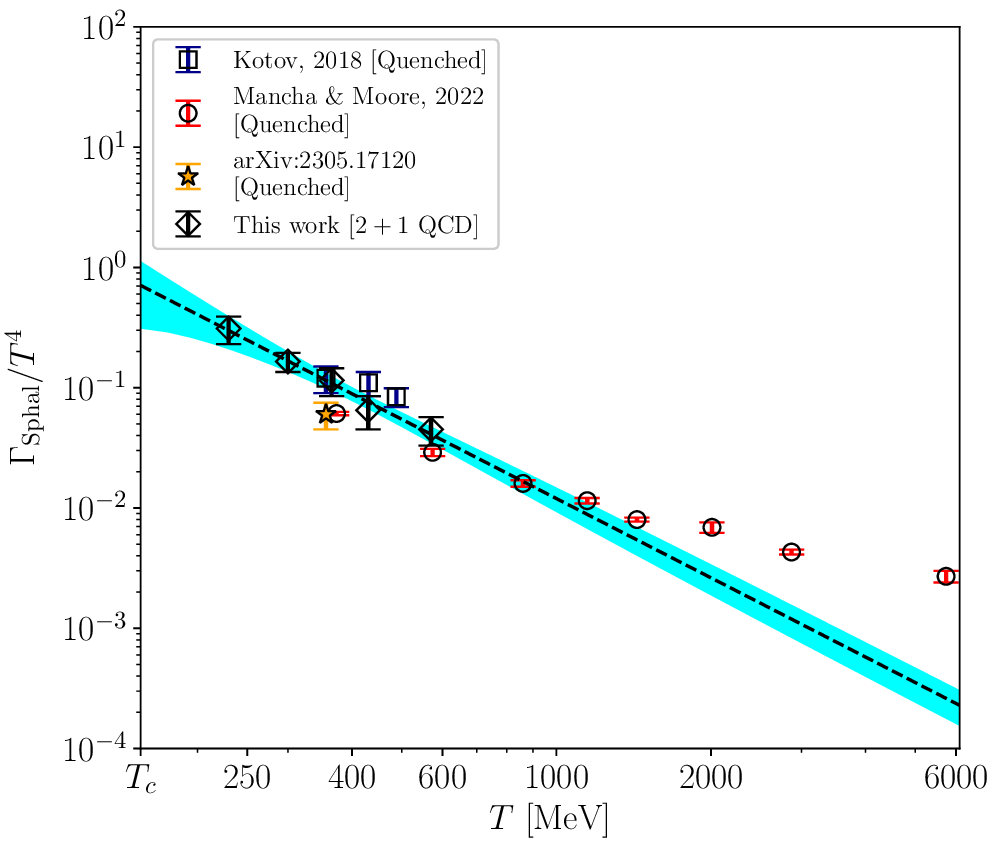}
\includegraphics[scale=0.47]{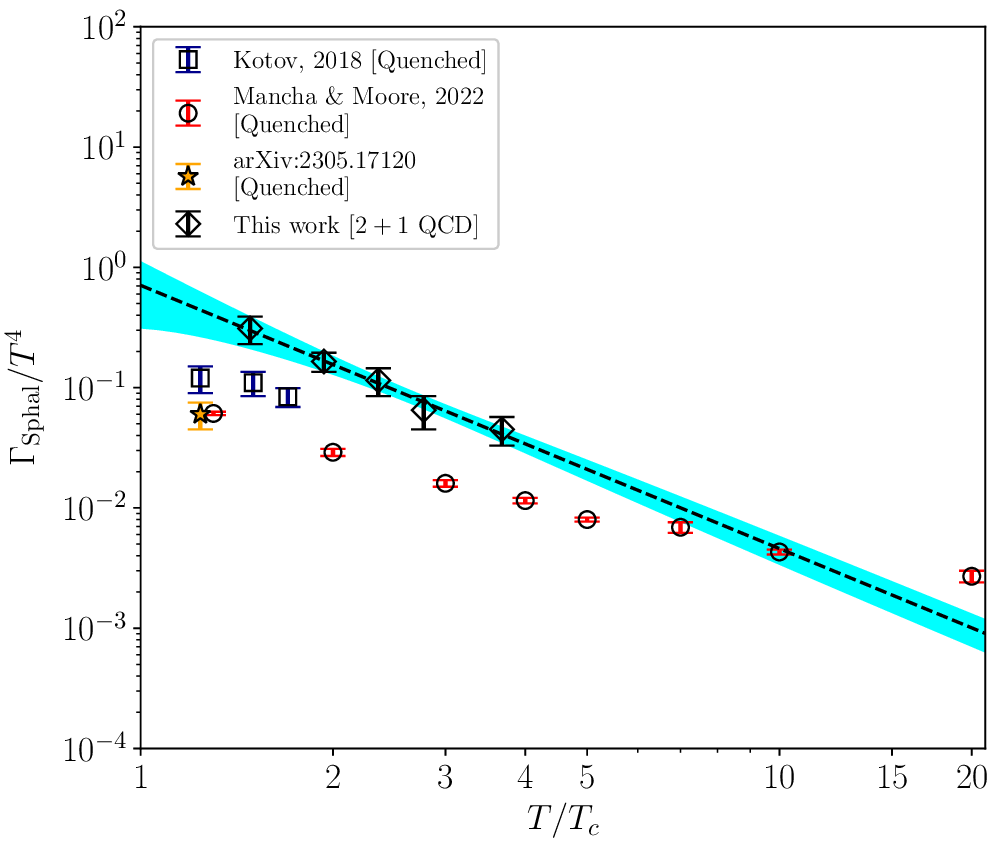}
\caption{Same as Fig.~\ref{fig:sphaleron_rate_fullQCD_ans1}, but using~\eqref{eq:fit_sphal_ans2} to fit our data.}
\label{fig:sphaleron_rate_fullQCD_ans2}
\end{figure}


As a final remark, we would also like to mention that, despite a semiclassically-inspired logarithmic power-law fits well our full QCD results for the sphaleron rate, also other functional forms could describe the $T$-behavior of our data. For example, a fit function of the type:
\beq\label{eq:fit_sphal_ans2}
\frac{\Gamma_\sphal}{T^4} = \widetilde{A} \left(\frac{T}{T_c}\right)^{-b},
\eeq
works perfectly fine as well, yielding a reduced chi-squared of 0.48/3, cf.~Fig.~\ref{fig:sphaleron_rate_fullQCD_ans2}, where the best fit with~\eqref{eq:fit_sphal_ans2} is depicted as a dashed line and a shaded area. Fit parameters turn out to be $\widetilde{A}=0.71(23)$ and $b=2.19(38)$.

\section{Conclusions}\label{sec:conclu}
In this letter we presented the first computation of the sphaleron rate in $2+1$ full QCD with physical quark masses as a function of the temperature in the range $200$~MeV~$\lesssim T \lesssim 600$~MeV.

The sphaleron rate was obtained from the inversion of finite lattice spacing and finite smoothing-radius lattice Euclidean topological charge density correlator from the modified Backus--Gilbert method recently introduced by the Rome group. Then, the physical value of the sphaleron rate was obtained performing a continuum limit at fixed smoothing radius, followed by a zero-smoothing limit.

Concerning the comparison of our full QCD determinations with previous quenched results, we found them to be larger. Concerning instead the temperature behavior of our data, our results for $\Gamma_\sphal/T^4$ can be fitted well by semiclassically-inspired functional form, predicting a logarithmic power-law decay of the rate. However, also other functional forms, such as a regular power-law decay of the rate, are shown to describe well our data.

Given that in this work we adopted a non-chiral discretization of the Dirac operator, the recovering of chiral symmetry in the continuum limit is a delicate point. Indeed, it is well known that the explicit breaking of the chiral symmetry of the staggered formulation leads to significant lattice artifacts in the topological susceptibility~\cite{Bonati:2015vqz,Petreczky:2016vrs,Borsanyi:2016ksw,Frison:2016vuc,Alexandrou:2017bzk,Bonati:2018blm,Athenodorou:2022aay}, that can be mainly traced back to a bad suppression of $Q\ne 0$ charge configurations in the path integral, due to the absence of exact zero-modes.

However, as shown in the Supplementary Material, thanks to the multicanonic algorithm we can easily compare correlators projected in a fixed topological sector with those obtained without projection. We observe, for our smallest temperature, where $\chi$ is less suppressed and thus the weight in the path integral of non-zero charge configurations is larger, that they perfectly agree within errors. This observation points out that a bad suppression of non-zero charge sectors, due to the absence of exact zero-modes, cannot be a significant source of lattice artifacts in the sphaleron rate computation. As a matter of fact, we find lattice artifacts for the sphaleron rate to be extremely mild (see Supplementary Material).

Finally, we stress that the same ensembles used here were also employed in~\cite{Athenodorou:2022aay} to compute the topological susceptibility at finite temperature, and for all temperatures the continuum limit of the gluonic discretization of $\chi$ was always confirmed by that obtained from a fermionic discretization based on the lowest-lying modes of the staggered operator~\cite{Bonanno:2019xhg}, which is affected by much smaller artifacts.

In conclusions, these observations make us confident that our continuum extrapolations for $\Gamma_\sphal$ are reliable. It would be extremely interesting to confirm our findings about the continuum scaling of $\Gamma_\sphal$ using a different fermionic discretization. Moreover, it would also be extremely interesting to repeat our calculation of the sphaleron rate, employing a different fermionic definition of the lattice topological charge density, based on the Index theorem, which is expected to suffer for smaller lattice artifact, or changing the pion mass, to explicitly check the behavior of $\Gamma_\sphal$ with $m_l$.

Another intriguing outlook would be to explore higher temperatures towards the GeV scale, in order to better clarify the actual temperature behavior of $\Gamma_\sphal$. At present, numerical limitations due to the infamous topological freezing problem~\cite{Alles:1996vn, DelDebbio:2004xh, Schaefer:2010hu, Luscher:2011kk, Bonati:2017woi} prevent us to reach higher temperatures, which would require to simulate very fine lattices with $a \lesssim 0.01$ fm, but possible algorithmic developments could permit such simulations in the next future~\cite{Hasenbusch:2017unr, Berni:2019bch, Bonanno:2018xtd,Bonanno:2020hht,Bonanno:2022yjr}.

Finally, it would be interesting to extend present computations to the case of non-zero spatial momentum $\vec{k}$. As a matter of fact, the momentum-dependence of the sphaleron rate is also of great phenomenological interest to compute the axion number density after decoupling using the Boltzmann equation~\cite{Notari:2022ffe}.

\section*{Acknowledgements}
It is a pleasure to thank G.~Gagliardi, V.~Lubicz, F.~Sanfilippo, G.~Villadoro and J.~H.~Weber for useful discussions. The work of C.~Bonanno is supported by the Spanish Research Agency (Agencia Estatal de Investigación) through the grant IFT Centro de Excelencia Severo Ochoa CEX2020-001007-S and, partially, by grant PID2021-127526NB-I00, both funded by MCIN/AEI/10.13039/501100011033. C.~Bonanno also acknowledges support from the project H2020-MSCAITN-2018-813942 (EuroPLEx) and the EU Horizon 2020 research and innovation programme, STRONG-2020 project, under grant agreement No 824093. Numerical simulations have been performed on the \texttt{MARCONI} and \texttt{Marconi100} machines at CINECA, based on the agreement between INFN and CINECA, under projects INF22\_npqcd and INF23\_npqcd.

\appendix
\section{SUPPLEMENTARY MATERIAL}

\section{Inversion method}

In this appendix we give more details about the inversion method proposed in~\cite{Hansen:2019idp} that we adopted to invert the relation in~\eqref{eq:rho_def} and obtain the spectral density.

For what follows, it is useful to rewrite Eq.~\eqref{eq:rho_def} as:
\beq
\label{eq:rho_with_Kprime}
G(t) = - \int_0^{\infty} \frac{d\omega}{\pi} \frac{\rho(\omega)}{\omega} K_t'(\omega),\\
\label{eq:Kprime}
K_t'(\omega) \equiv \omega \frac{\cosh\left[\omega(t-1/(2T))\right]}{\sinh\left[\omega/(2T)\right]}.
\eeq
From the combination of Eqs.~\eqref{eq:rho_with_Kprime} and~\eqref{eq:Estimator}, it is possible to relate the spectral density estimator $\bar{\rho}(\bar{\omega})$ and the physical spectral function $\rho(\omega)$ via the resolution function $\Delta(\omega,\bar{\omega})$:
\beq\label{eq:rel_smeared}
\frac{\bar{\rho}(\bar{\omega})}{\bar{\omega}} = \int_0^{\infty} d\omega \Delta(\omega,\bar{\omega}) \frac{\rho(\omega)}{\omega},
\eeq
where 
\beq\label{eq:resolution}
\Delta(\omega,\bar{\omega})= \sum_{t=0}^{1/T} g_t(\bar{\omega}) K'_t(\omega).
\eeq
To obtain a faithful reconstruction of the physical spectral density from the estimator $\bar{\rho}$, it is necessary that the resolution function is sufficiently peaked around $\bar{\omega}$ as a function of $\omega$. Therefore, constraining the shape of the resolution function (i.e., the value of the coefficients $g_t$), is a crucial step to determine the quality of our inversion.

The modified Backus--Gilbert method proposed in Ref.~\cite{Hansen:2019idp} consists in determining the $g_t$ coefficients from the minimization of the following functional:
\beq\label{eq:backus_gilbert_functional_tot}
F[g_t] = (1-\lambda) A_\alpha[g_t] +  \frac{\lambda}{\mathcal{C}^2}B[g_t], \,\quad  \lambda\in [0,1).
\eeq
Here $\mathcal{C}$ is a normalization factor proportional to the correlator in a fixed point (we used $\mathcal{C}=G(tT=0.5)$ in this work) and $\lambda$ is a free parameter which is varied to check for systematics; instead, the functionals $A_\alpha$ and $B$ are defined as follows:
\beq
A_\alpha[g_t] = \int_0^{\infty} d\omega \, [\Delta(\omega,\bar{\omega}) - \delta(\omega,\bar{\omega})]^2 \,e^{\alpha \omega},
\eeq
with $\alpha < 2$, and
\beq
B[g_t] = \sum_{t,t'=0}^{1/T} \Cov_{t,t'} \, g_t g_{t'}.
\eeq
The first functional minimizes the distance between the resolution function and a given target function $\delta(\omega,\bar{\omega})$, which is chosen on the basis of physical considerations. An exponentially-growing factor is added in the integral to promote asymptotically large frequencies~\cite{ExtendedTwistedMassCollaborationETMC:2022sta}, as $\rho(\omega)$ is predicted to grow as a power-law of $\omega$ from perturbation theory~\cite{Laine:2011xm} (here we used $\alpha=1.99$). The second functional depends instead on the covariance matrix $\Cov_{t,t'} = \braket{[G(t)-\braket{G(t)}][G(t')-\braket{G(t')}]}$ of the correlator $G(t)$ and is used to regularize the inversion problem. We stress here that the reconstruction of the sphaleron rate was done using all available $t>0$ points of the correlator $G(t)$. Although the short-distance behavior of the correlator is affected by cooling, we find that the inclusion/exclusion of the first few points of the correlator from the Backus--Gilbert minimization did not have a sizable impact on the results for the sphaleron rate. Such behavior is expected, and is due to the fact that the sphaleron rate is rather insensitive to the short-distance behavior of $G(t)$ (see discussion below).

Following the lines of Ref.~\cite{Almirante:2023wmt,Bonanno:2023ljc}, also here we used the pseudo-Gaussian target function, which is inspired by the shape of $K_t'(\omega)$ for the largest time separation $t=1/(2T)$, cf.~Eq.~\eqref{eq:Kprime}:
\beq\label{eq:target}
\delta(\omega, \bar{\omega}=0) = \left(\frac{2}{\sigma \pi}\right)^2 \frac{\omega}{\sinh(\omega/\sigma)}.
\eeq
The parameter $\sigma$, which determines the width of the target function, is crucial to obtain a reliable result for the $g_t$ coefficients. Ideally, one would like to have $\sigma \to 0$, as a peaked target function reflects in a peaked resolution function, leading to a better approximation of $\rho$ via $\bar{\rho}$ (as a matter of fact, in this limit the smearing function tends to a Dirac delta). However, statistical errors on $g_t$ (and thus on the sphaleron rate) tend to explode when $\sigma$ is decreased. This is due to the fact the the inversion problem is not a well-defined mathematical problem. On the other hand, if $\sigma$ is increased, statistical errors decrease and the signal for $g_t$ stabilizes, but at the same time systematic effects introduced by the finiteness of the width of the peak of the resolution function become dominant, and our quantity $\bar{\rho}$ becomes less and less reliable as an estimator of the physical spectral density. Therefore, the choice of the parameter $\sigma$ has to be done with some care, trying to find a good compromise between the two facts.

In this work, from the study of the dependence on $\sigma$ of our results, we found that choosing $\sigma/T=1.75$, and keeping it fixed in physical units for all ensembles at different temperatures, is a reasonable choice. As a matter of fact, from general theoretical arguments, being our target function even in $\omega/\sigma$, we expect, for asymptotically small smearing widths~\cite{Bulava:2021fre,Frezzotti:2023nun,Evangelista:2023vtl}:
\beq\label{eq:zerosigma_extr}
\Gamma_\sphal(\sigma) = \Gamma_\sphal(0) + c (\sigma/T)^2 + O[(\sigma/T)^4].
\eeq
Thus, since this expansion starts from $O(\sigma^2)$, we expect to observe a mild dependence on $\sigma$ if it is chosen small enough. We checked that indeed $\sigma/T=1.75$ is a choice that works well in this case. More details are discussed in the next section of the Supplementary Material.

Finally, the role of the $\lambda$ parameter is to look for systematic effects in the determination of the sphaleron rate from the Backus--Gilbert method. This is necessary as the determination of the spectral density as an inverse problem is not a well-defined mathematical problem. To look for systematics, we study the sphaleron rate as a function of the quantity:
\beq\label{eq:def_d}
d[g_t](\lambda) \equiv \sqrt{\frac{A_0[g_t]}{B[g_t]}}.
\eeq
When systematic effects are under control $d[g_t]\ll 1$ and the rate does not depend on $d[g_t]$ within the errors. Thus, we look for a window in $\lambda$ where $d[g_t](\lambda)\ll 1$, where errors are statistically dominated, and we include in our estimations any small systematic variation observed in this window. For more details about this, we refer the reader to the extensive discussion of Ref.~\cite{Bonanno:2023ljc} about this point, as here we follow exactly the same strategy.

\section{Additional plots}

\begin{figure*}[!htbp]
\includegraphics[scale=0.47]{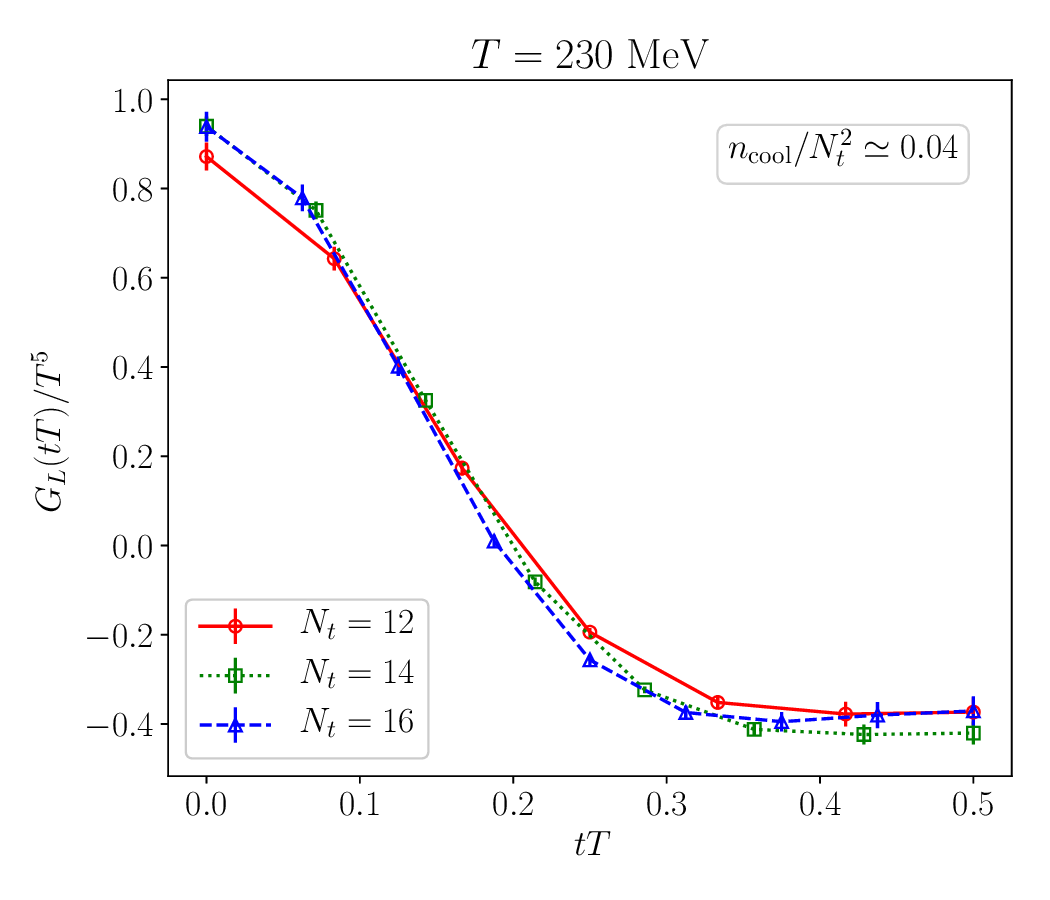}
\includegraphics[scale=0.48]{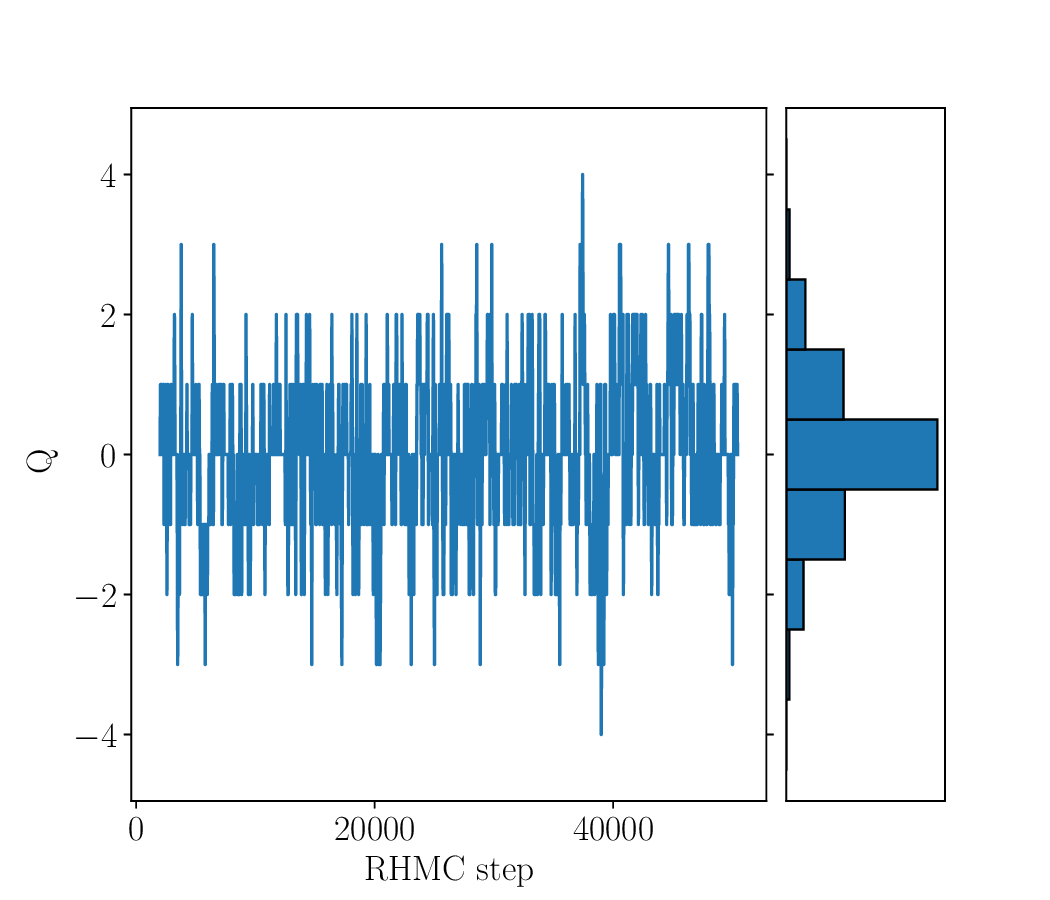}
\includegraphics[scale=0.49]{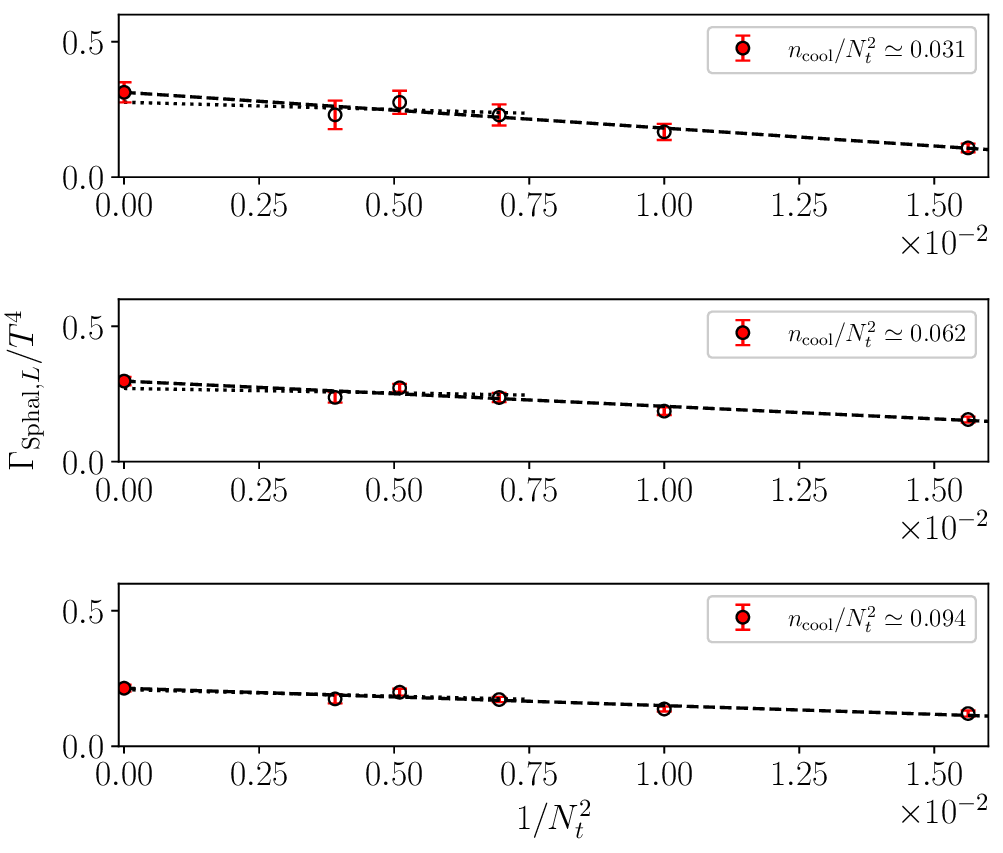}
\includegraphics[scale=0.5]{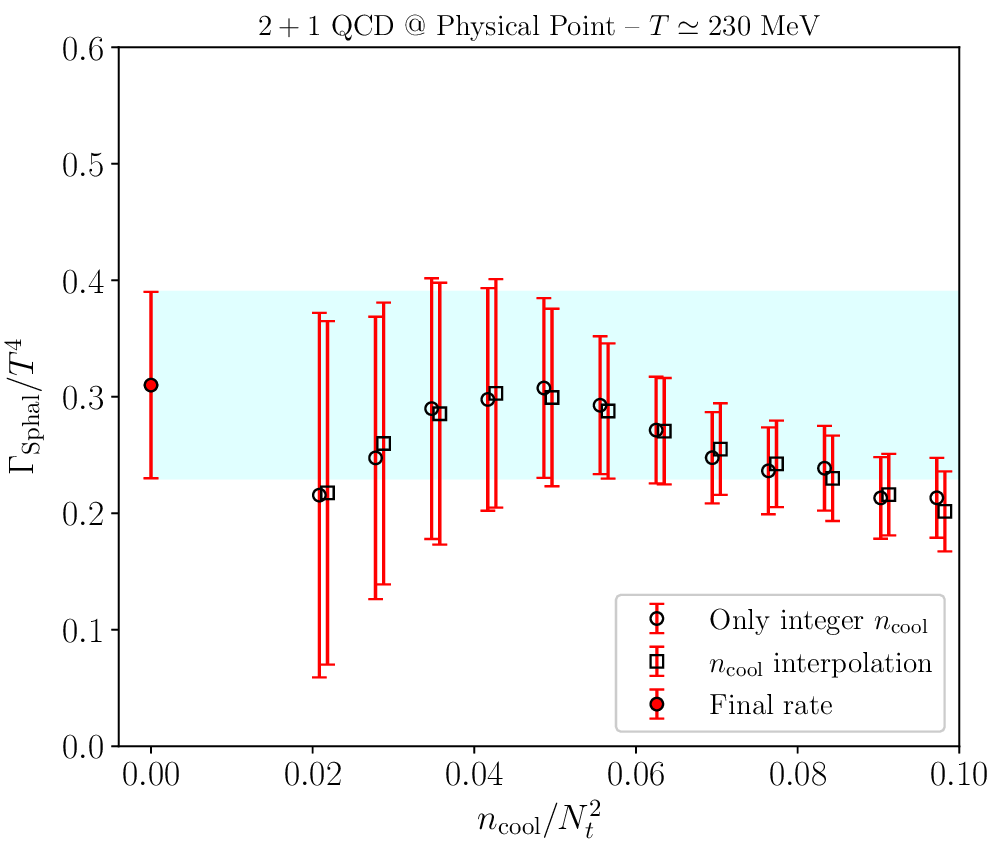}
\includegraphics[scale=0.57]{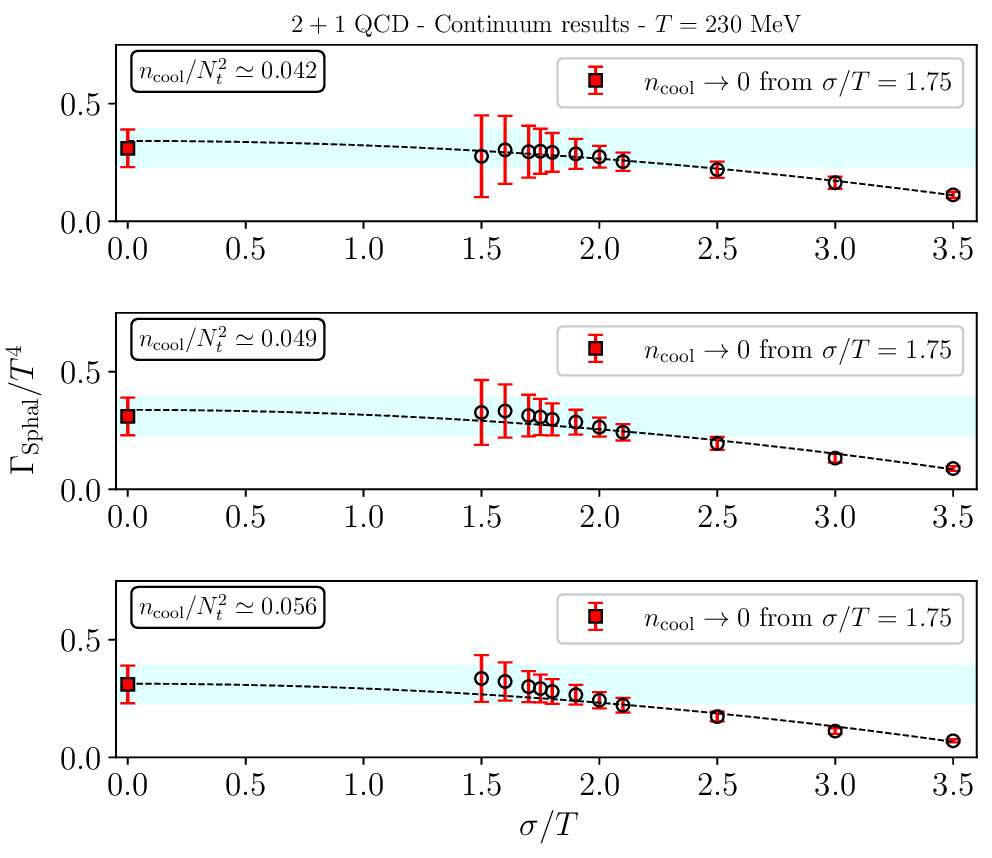}
\caption{Plots in this figure refer to $T=230~\mathrm{MeV}$: comparison of the topological charge density correlators for the three finest lattice spacings, at approximately the same value of $n_\cool/N_t^2$ (top left); history of the Monte Carlo evolution (and corresponding distribution) of the topological charge $Q$ for the finest lattice spacing obtained using the multicanonic algorithm (top right); examples of the continuum limit extrapolation of $\Gamma_{\sphal,L}/T^4$ at fixed $n_\cool/N_t^2$ (center left); continuum limit of $\Gamma_\sphal/T^4$ as a function of $n_\cool/N_t^2$, for $\sigma/T=1.75$, together with a zero smoothing-radius extrapolation (center right); continuum limit of $\Gamma_\sphal/T^4$, as a function of $\sigma/T$, for few values of $n_\cool/N_t^2$, the continuous lines represent the result of a fit assuming $O(\sigma^2)$ corrections, which are compared in $\sigma = 0$ with our final estimate, to show that systematic effects are well under control (bottom).}
\label{fig:T230}
\end{figure*}

\begin{figure*}[!htbp]
\includegraphics[scale=0.47]{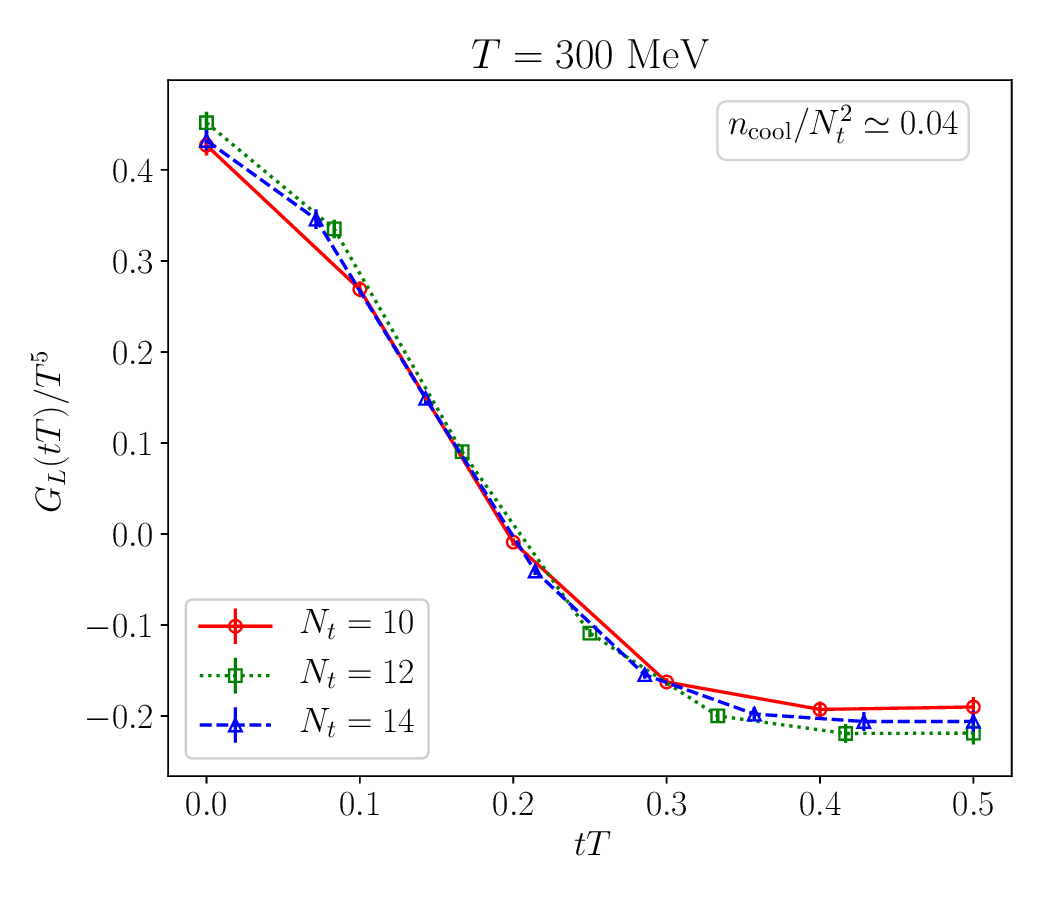}
\includegraphics[scale=0.48]{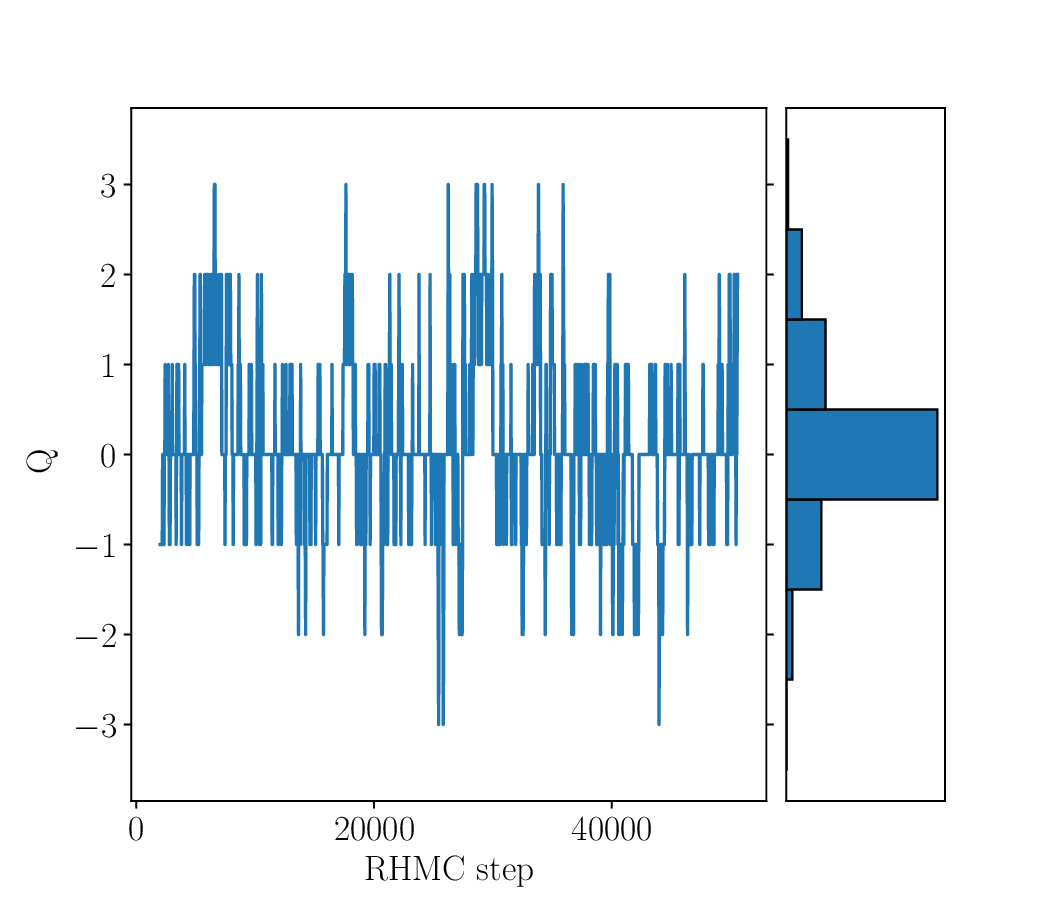}\hspace{20cm}
\includegraphics[scale=0.49]{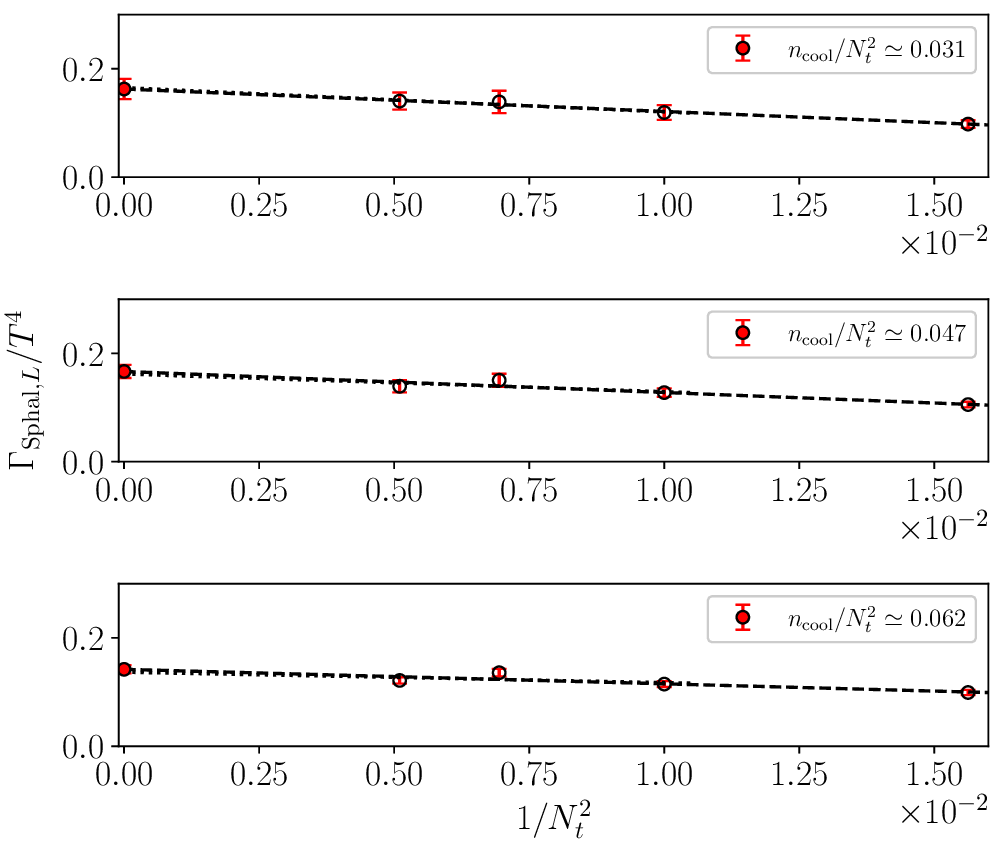}
\includegraphics[scale=0.50]{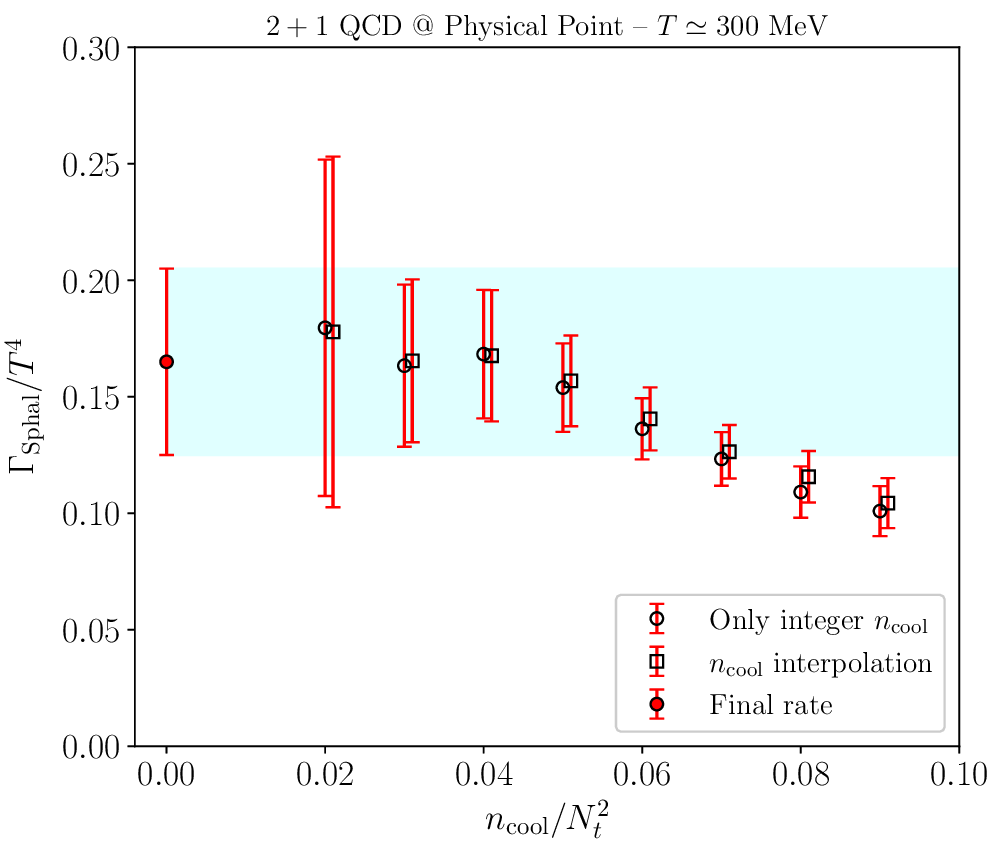}
\includegraphics[scale=0.57]{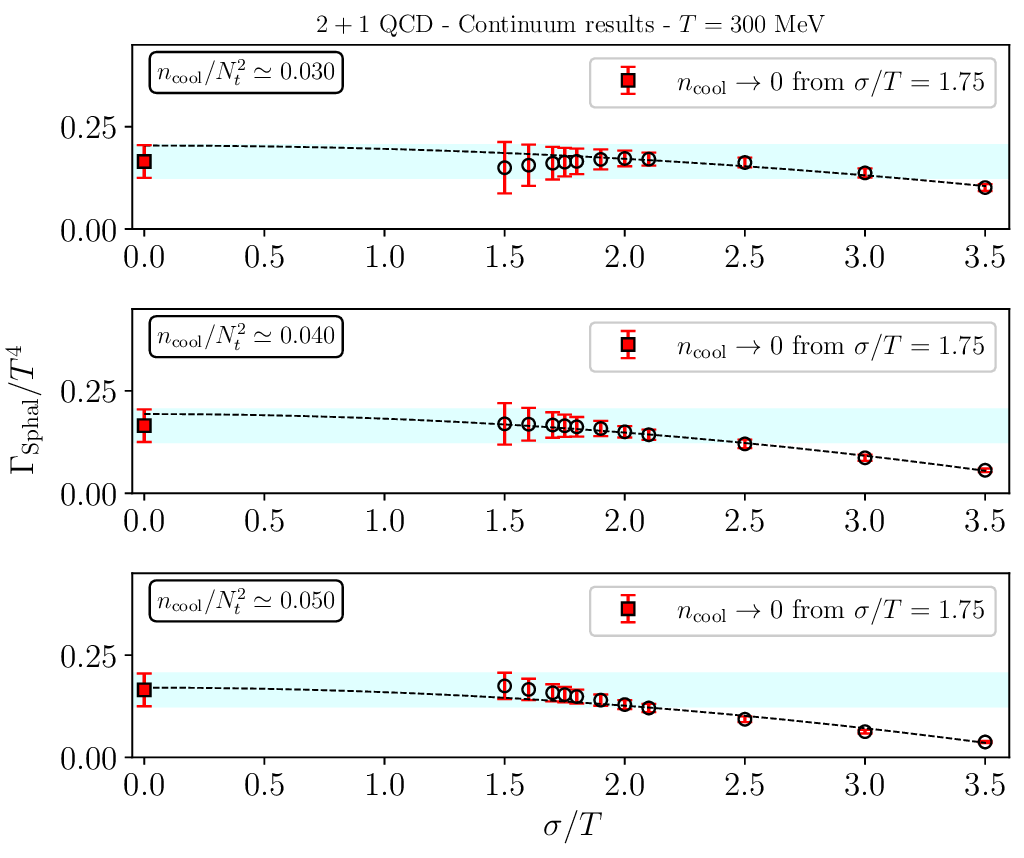}
\caption{Same as in Fig.~\ref{fig:T230}, for a temperature $T=300$~MeV.}
\label{fig:T300}
\end{figure*}

\begin{figure*}[!htbp]
\includegraphics[scale=0.47]{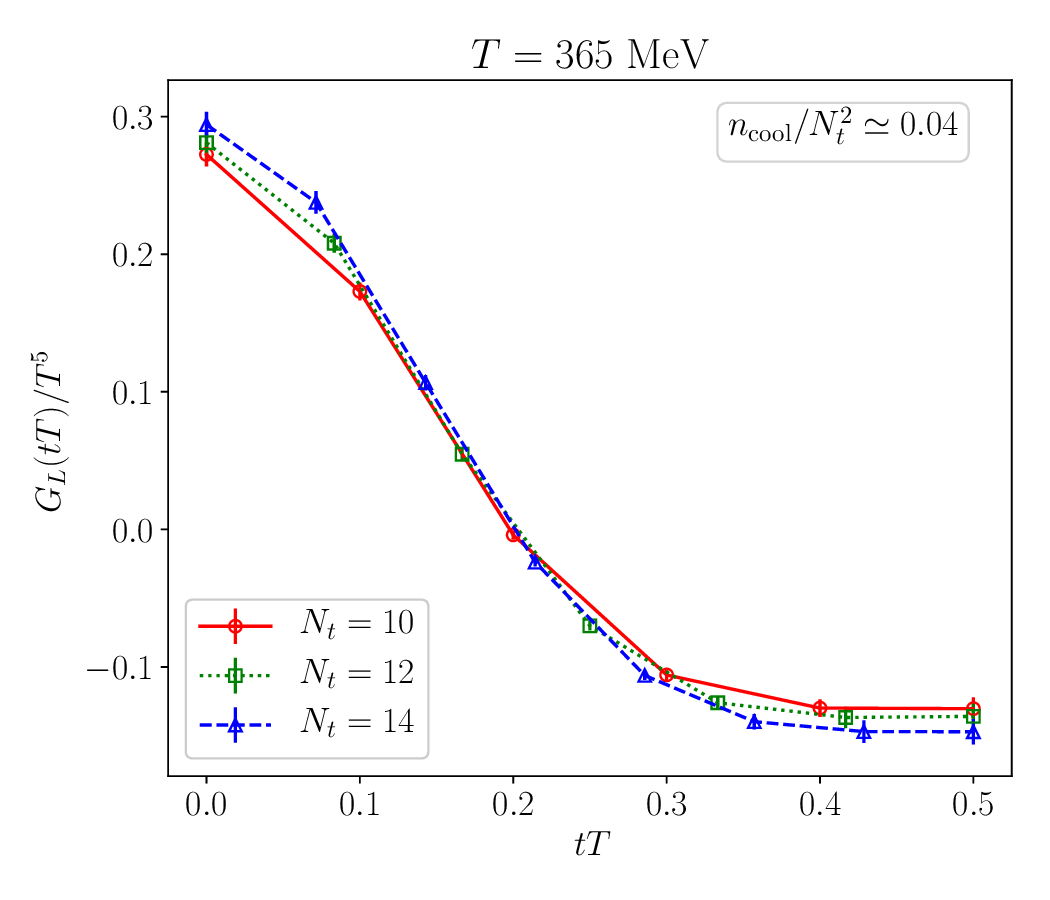}
\includegraphics[scale=0.48]{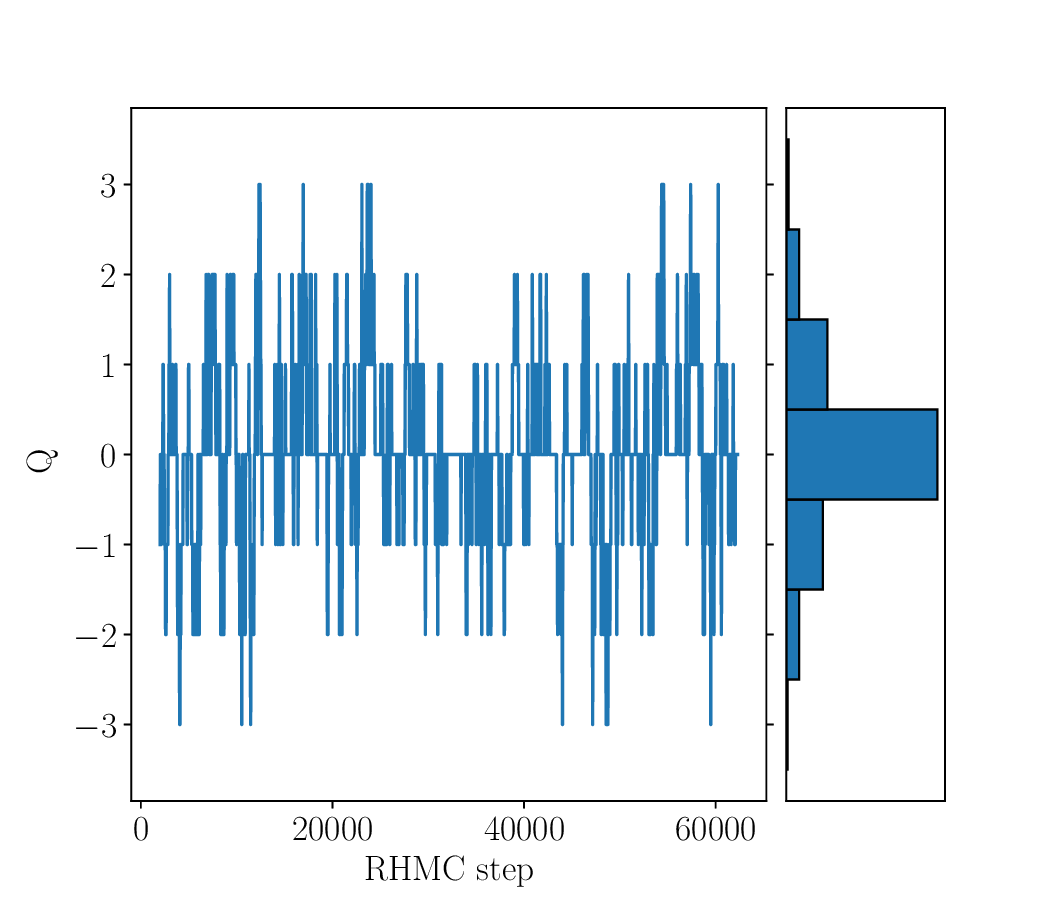}\hspace{20cm}
\includegraphics[scale=0.49]{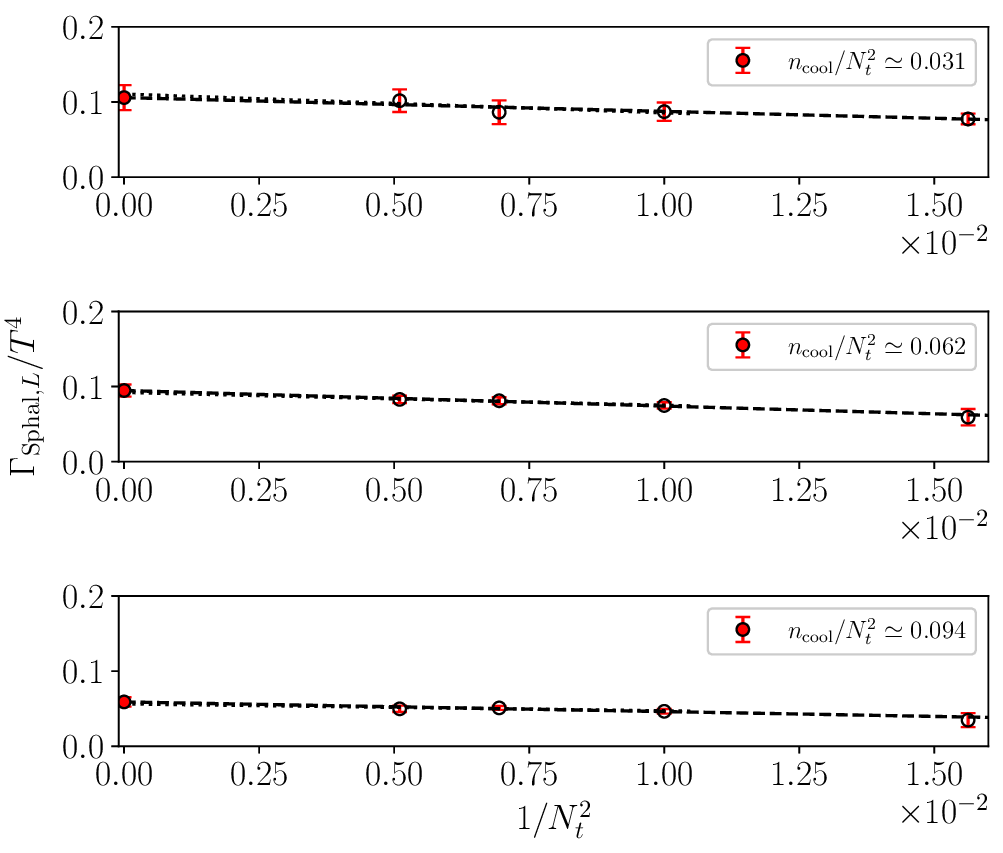}
\includegraphics[scale=0.5]{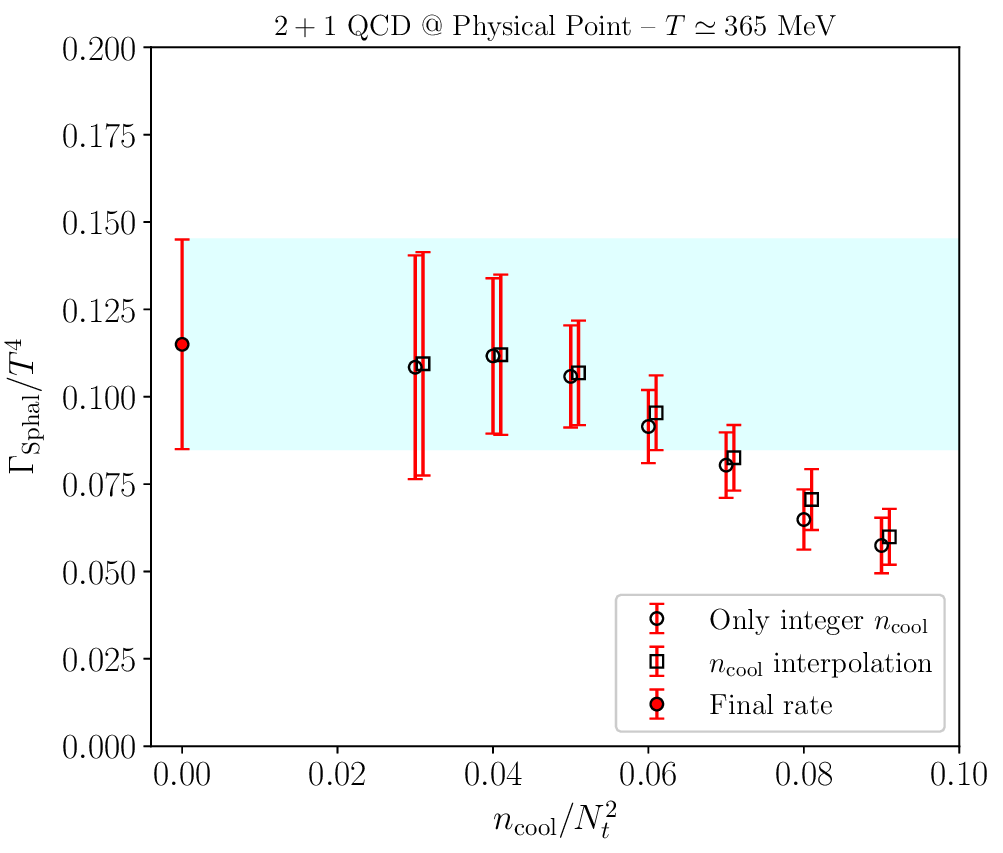}
\includegraphics[scale=0.57]{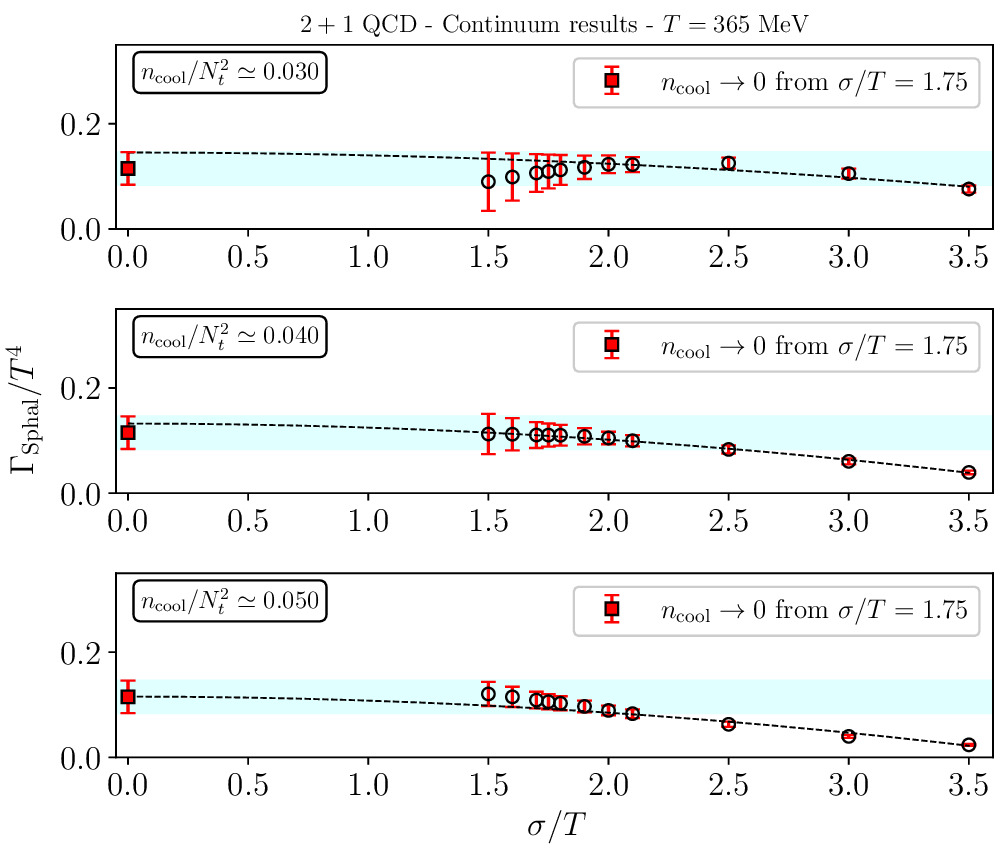}
\caption{Same as in Fig.~\ref{fig:T230}, for a temperature $T=365$~MeV.}
\label{fig:T365}
\end{figure*}

\begin{figure*}[!htbp]
\includegraphics[scale=0.47]{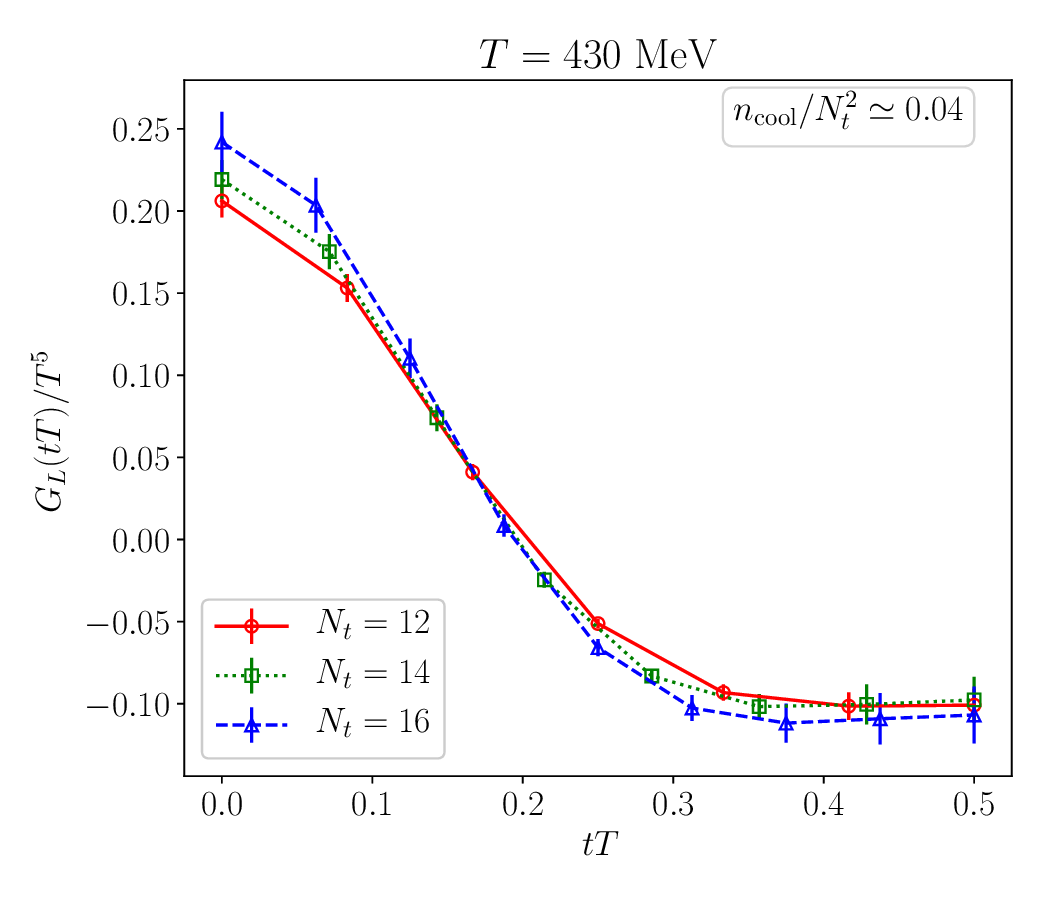}
\includegraphics[scale=0.48]{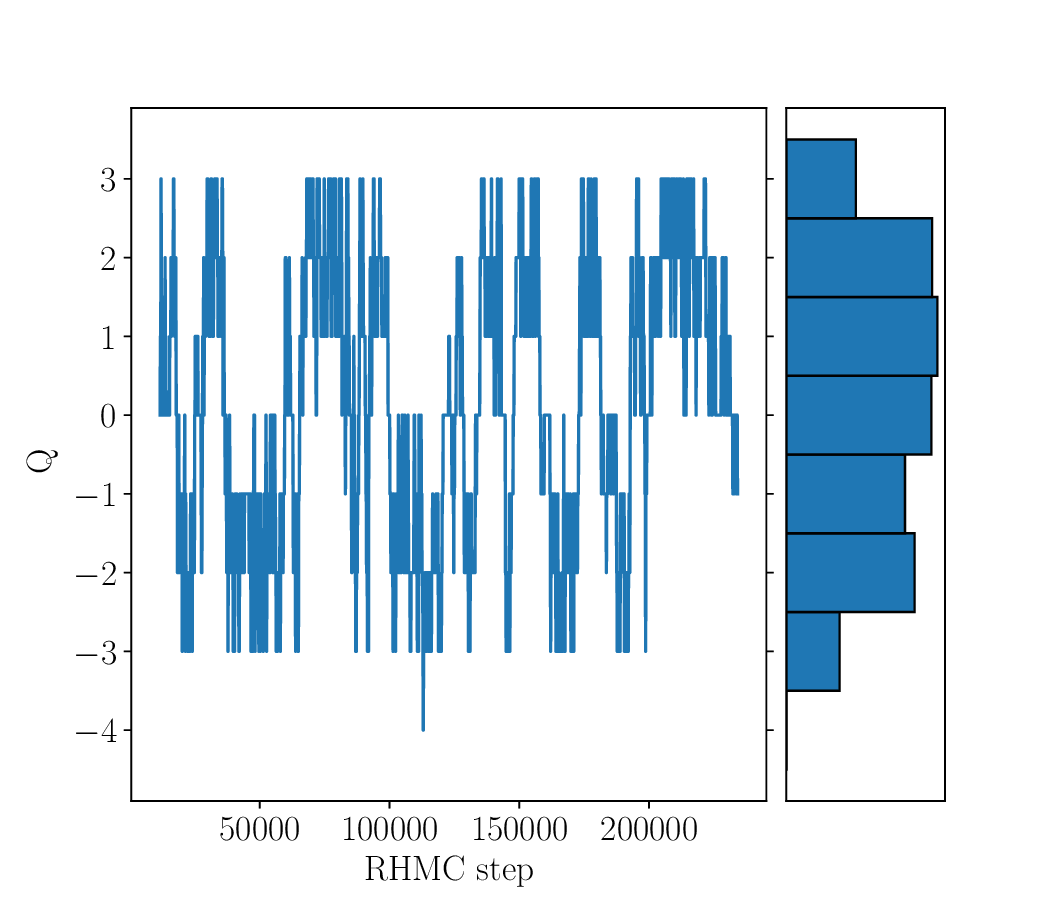}\hspace{20cm}
\includegraphics[scale=0.49]{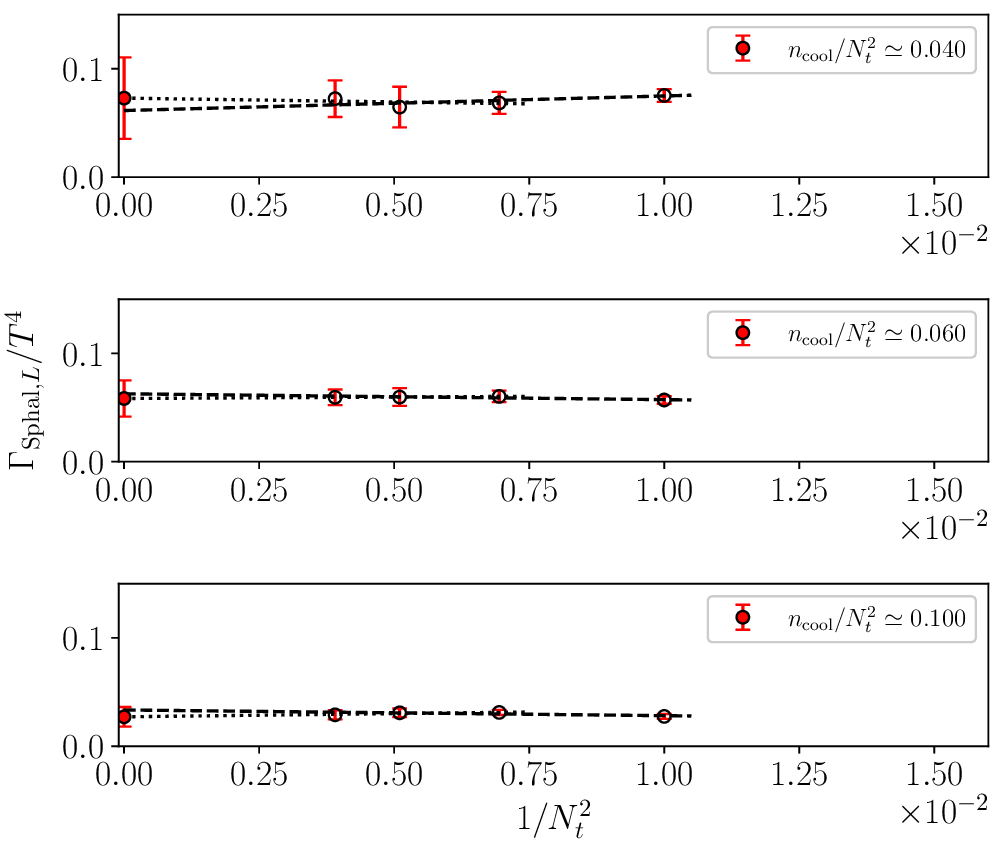}
\includegraphics[scale=0.5]{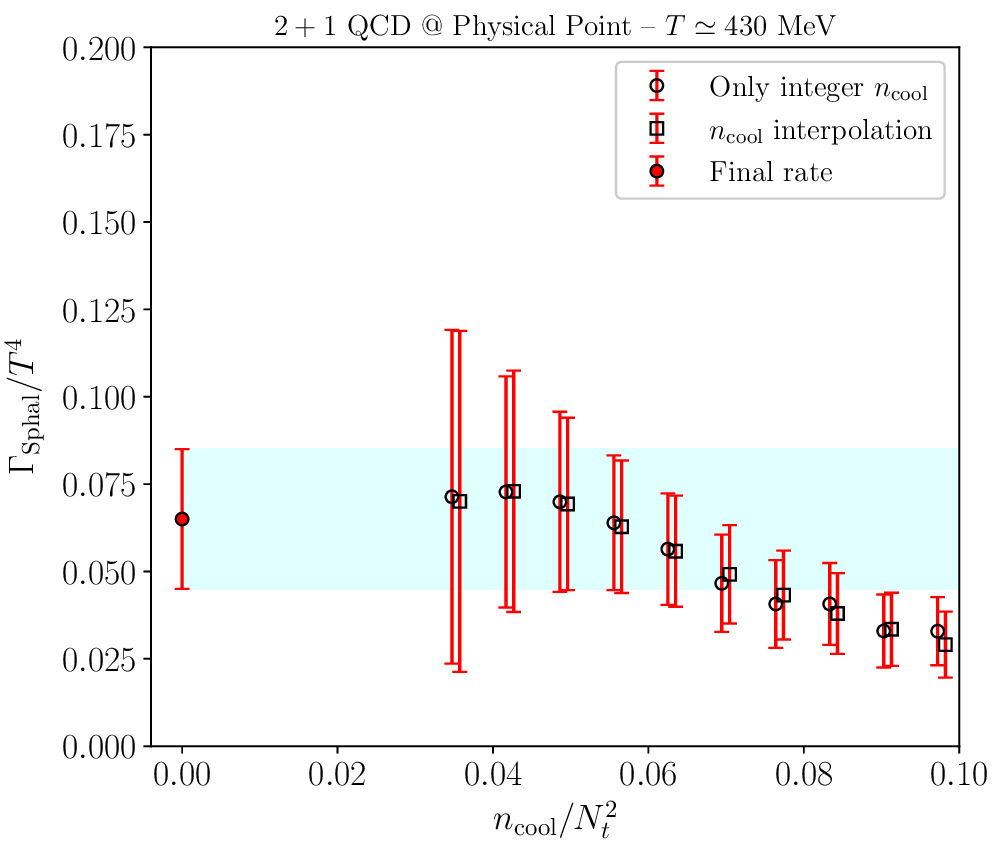}
\includegraphics[scale=0.57]{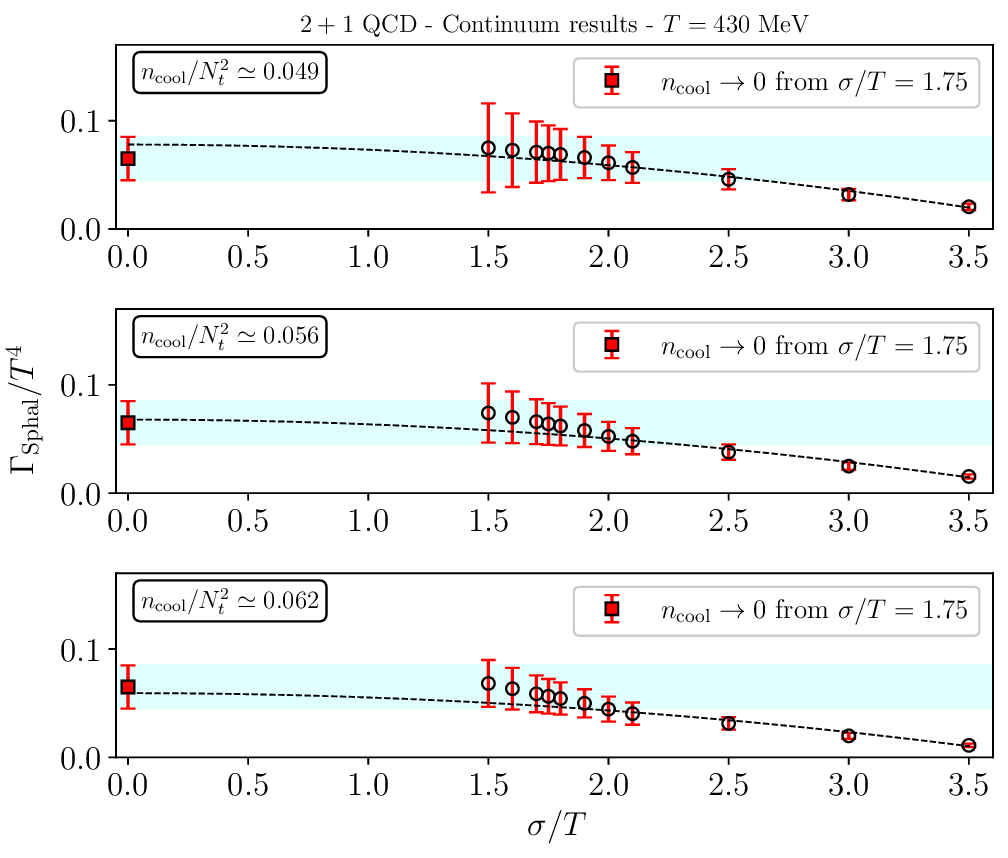}
\caption{Same as in Fig.~\ref{fig:T230}, for a temperature $T=430$~MeV.}
\label{fig:T430}
\end{figure*}

\begin{figure*}[!htbp]
\includegraphics[scale=0.47]{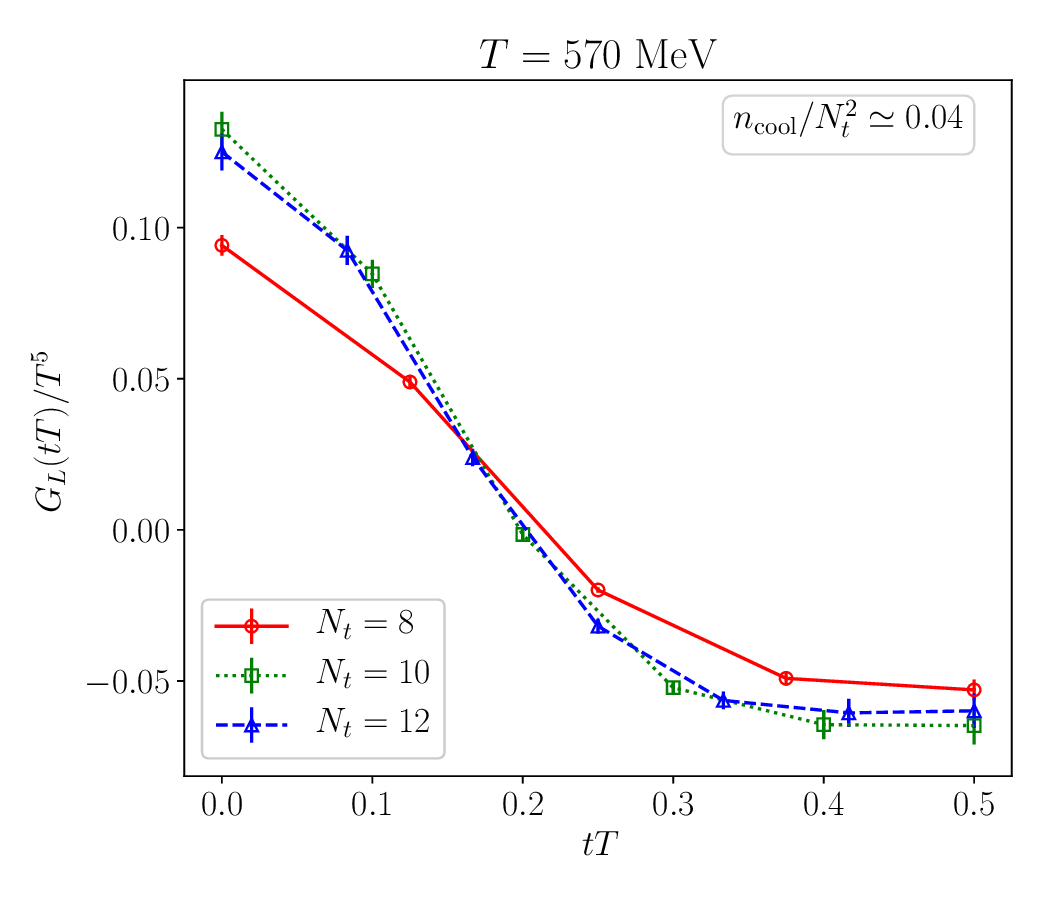}
\includegraphics[scale=0.48]{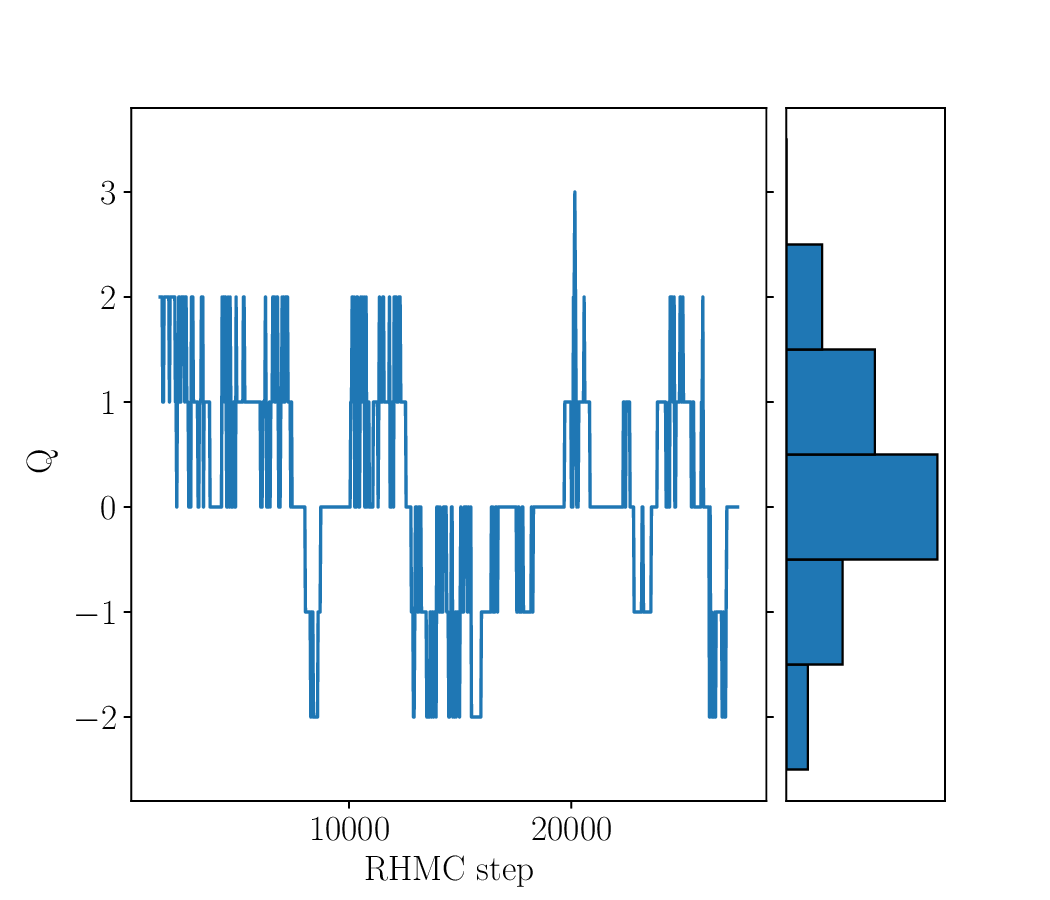}\hspace{20cm}
\includegraphics[scale=0.49]{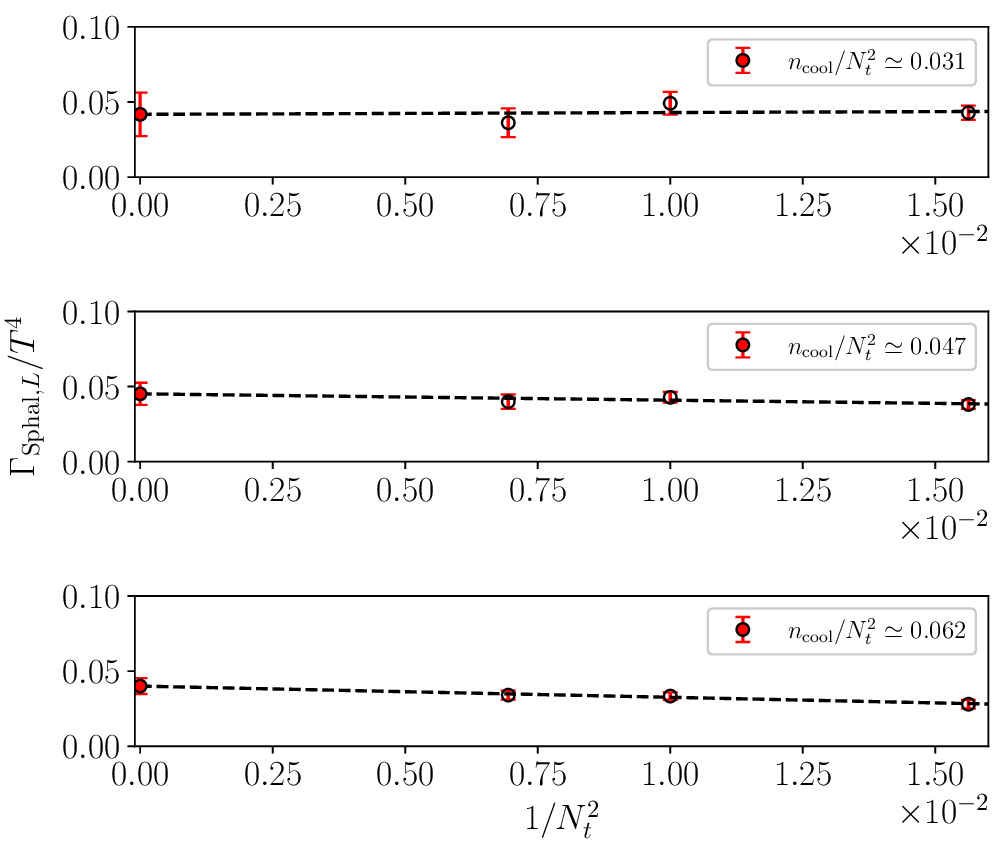}
\includegraphics[scale=0.5]{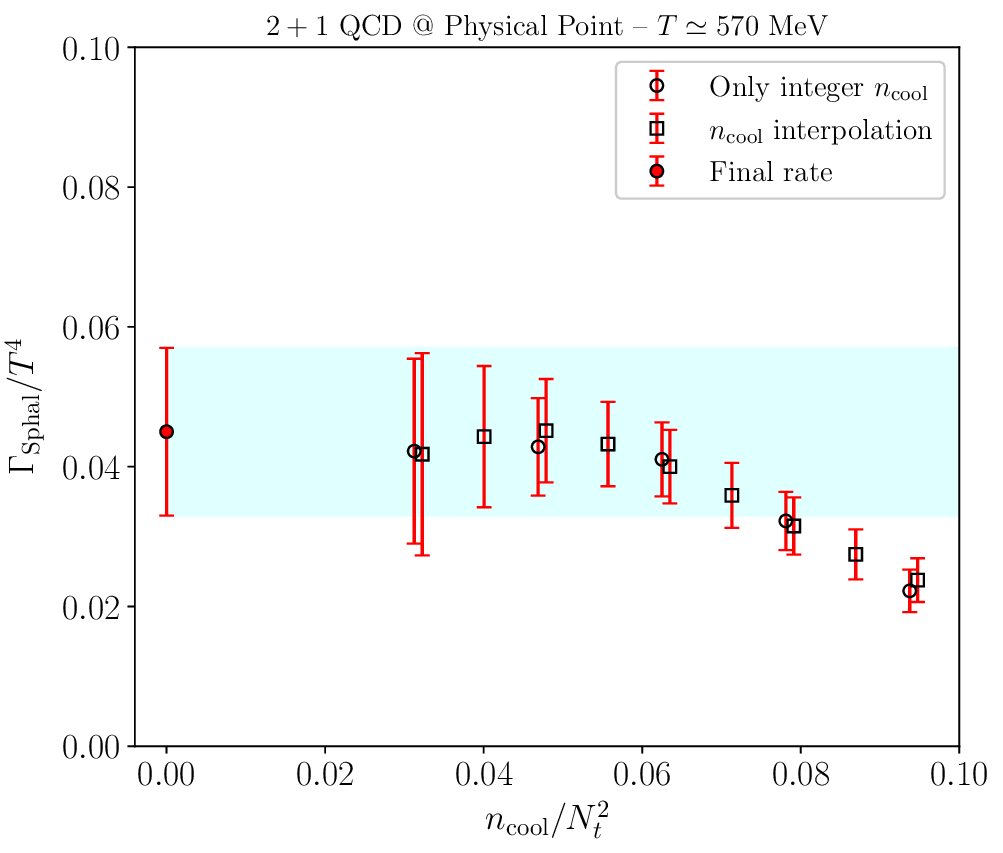}
\includegraphics[scale=0.57]{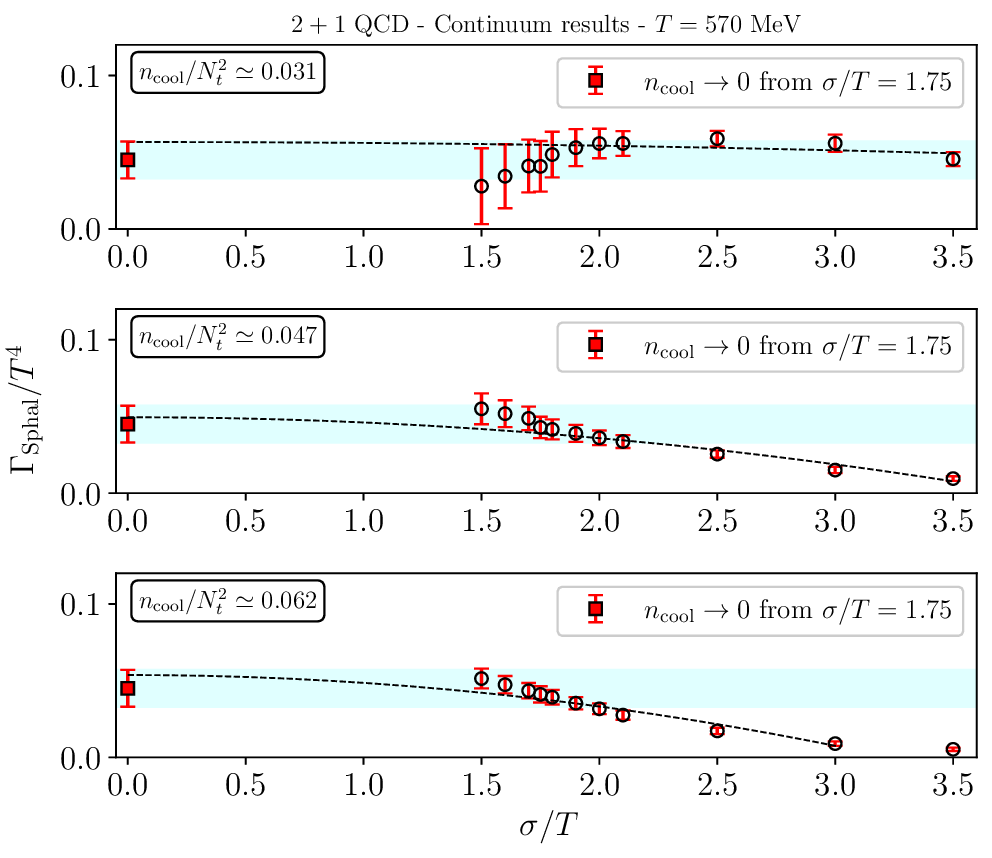}
\caption{Same as in Fig.~\ref{fig:T230}, for a temperature $T=570$~MeV.}
\label{fig:T570}
\end{figure*}

In Figs.~\ref{fig:T230},~\ref{fig:T300},~\ref{fig:T365},~\ref{fig:T430},~\ref{fig:T570} we collect the following additional plots for each explored temperature:
\begin{itemize}
\item The Monte Carlo evolution of the topological charge $Q$ as a function of the multicanonic RHMC step for the finest lattice spacing available for that temperature, and its related histogram. All shown histories are obtained before reweighting, i.e., in the presence of the bias topological potential. We defined the integer topological charge as $Q = \mathrm{round}[\alpha_0 Q_\cool]$, where $\mathrm{round}(x)$ denotes the closest integer to $x$, $Q_\cool$ is the clover lattice topological charge in Eqs.~\eqref{eq:topcharge_dens_clover},~\eqref{eq:topcharge_profile} computed after cooling and $\alpha_0$ is a parameter found minimizing the mean square distance between $\alpha Q_\cool$ and $\mathrm{round}[\alpha Q_\cool]$. Since the topological susceptibility exhibited a plateau after $n_\cool\approx 80$ in all cases, we chose $n_\cool = 100$ for each ensemble to compute $Q$.
\item The behavior of the Euclidean lattice topological charge density correlators $G_L(tT)$ for the three available finest lattice spacings, computed at approximately the same value of $n_\cool/N_t^2$.
\item Continuum extrapolations of $\Gamma_{\sphal,L}$ at fixed smoothing radius for a few values of $n_\cool/N_t^2$, performed assuming $O(a^2) = O(1/N_t^2)$ corrections to the continuum limit. In all cases, we took the continuum extrapolation obtained fitting the three finest lattice spacings (dotted line). However, we verified that a linear fit in $1/N_t^2$ including all available points always gave compatible results within errors (dashed line). To keep $n_\cool/N_t^2$ fixed, one can follow two strategies: one consists in doing a spline interpolation of $\Gamma_{\sphal,L}$ as a function of $n_\cool$ to define it for non-integer cooling steps. Then, $n_\cool/N_t^2$ can be kept fixed for each lattice spacing. The second is to use only integer values using the following approximation: $n_\cool' = \mathrm{round} \left[ n_\cool (N_t'/N_t)^2 \right]$. In this case, $n_\cool/N_t^2$ can be kept constant among different ensembles with different lattice spacings only up to $\sim3-4\%$ differences. In all cases, we observed no difference in the obtained continuum extrapolations, as already observed in the quenched case~\cite{Bonanno:2023ljc}. This is due to the fact that the dependence of $\Gamma_{\sphal,L}$ on $n_\cool$ is extremely mild, especially for smaller values of the number of cooling steps. This behavior has a physical explanation: the sphaleron rate is the zero-frequency limit of $\rho(\omega)$, thus it is highly insensible to the UV scale set by the smoothing radius (see the extended discussion of~\cite{Bonanno:2023ljc} on this point).
\item Behavior of the continuum extrapolations of $\Gamma_\sphal$ as a function of $n_\cool/N_t^2$ for $\sigma/T=1.75$ and our final result for the sphaleron rate, depicted as a full round point and as a shaded area. In this plot we show both the continuum extrapolations at fixed value of $n_\cool/N_t^2$ obtained by using only integer values of $n_\cool$ (as explained earlier) and the extrapolations obtained by interpolating finite-lattice-spacing determinations of the sphaleron rate as a function of $n_\cool$, using a spline interpolation. In all cases we observe perfect agreement between these two different approximations. Since, as expected on the basis of physical arguments and from the definition of $\Gamma_\sphal$, the sphaleron rate exhibits a plateau as a function of the smoothing radius for small enough values of $n_\cool/N_t^2$, we do not perform any zero-cooling extrapolation, and simply take the value at the plateau as our final determination. The central values and the errors, represented as a full point in $n_\cool/N_t^2=0$ and a shaded area, were assessed on the basis of the central values and the errors of the points within the plateau.
\item Behavior of the continuum extrapolations of $\Gamma_\sphal$ as a function of the smearing width $\sigma/T$ for a few values of $n_\cool/N_t^2$, chosen to stay within the plateau observed for $\sigma/T=1.75$. In all cases, we observe that our data approach a plateau in the range $1.5 \le \sigma/T \le 2$, thus justifying our choice $\sigma/T = 1.75$ to compute the sphaleron rate. As a further confirmation, we observe that our data are very well described by a functional form of the type $A_1 + A_2 (\sigma/T)^2$ in the range $\sigma/T \in [1.5,3.5]$, as expected from general theoretical arguments, cf.~Eq.~\eqref{eq:zerosigma_extr}. Performing a best fit of our results to that functional form, the fit parameter $A_1$ gives an estimate of the $\sigma \to 0$ extrapolation. In all cases we find that such extrapolation agrees with our final determination for the sphaleron rate, obtained from the behavior of the rate as a function of $n_\cool/N_t^2$ at $\sigma/T=1.75$. Thus, we just took the determinations obtained at $\sigma/T=1.75$ as our final results.
\end{itemize}

\begin{figure}[!t]
\centering
\includegraphics[scale=0.49]{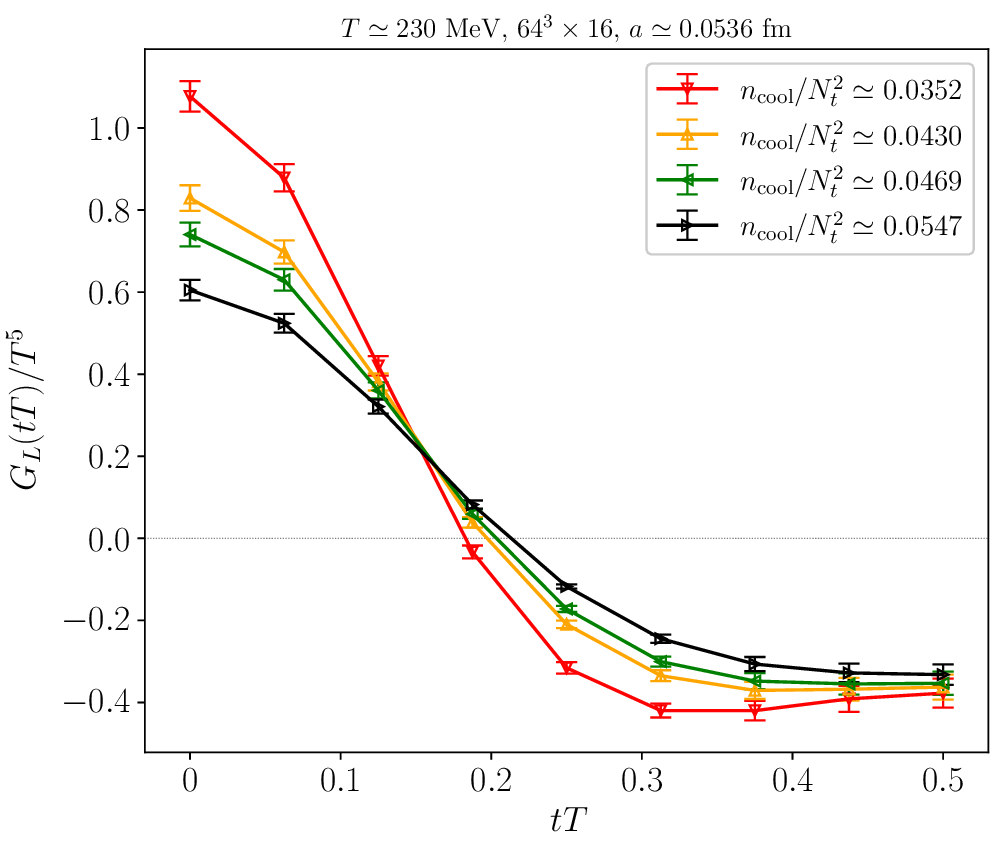}
\caption{Topological charge density correlator at $T=230\mathrm{MeV}$ for our finest lattice spacing ($N_t=16$) as a function of the smoothing radius. Shown values of the number of cooling steps have been chosen so as to stay within the plateau observed for $\Gamma_\sphal$ as a function of $n_\cool$, cf.~Fig.~\ref{fig:T230}.}
\label{fig:tcorr_comp_ncool}
\end{figure}

\begin{figure}[!t]
\centering
\includegraphics[scale=0.4]{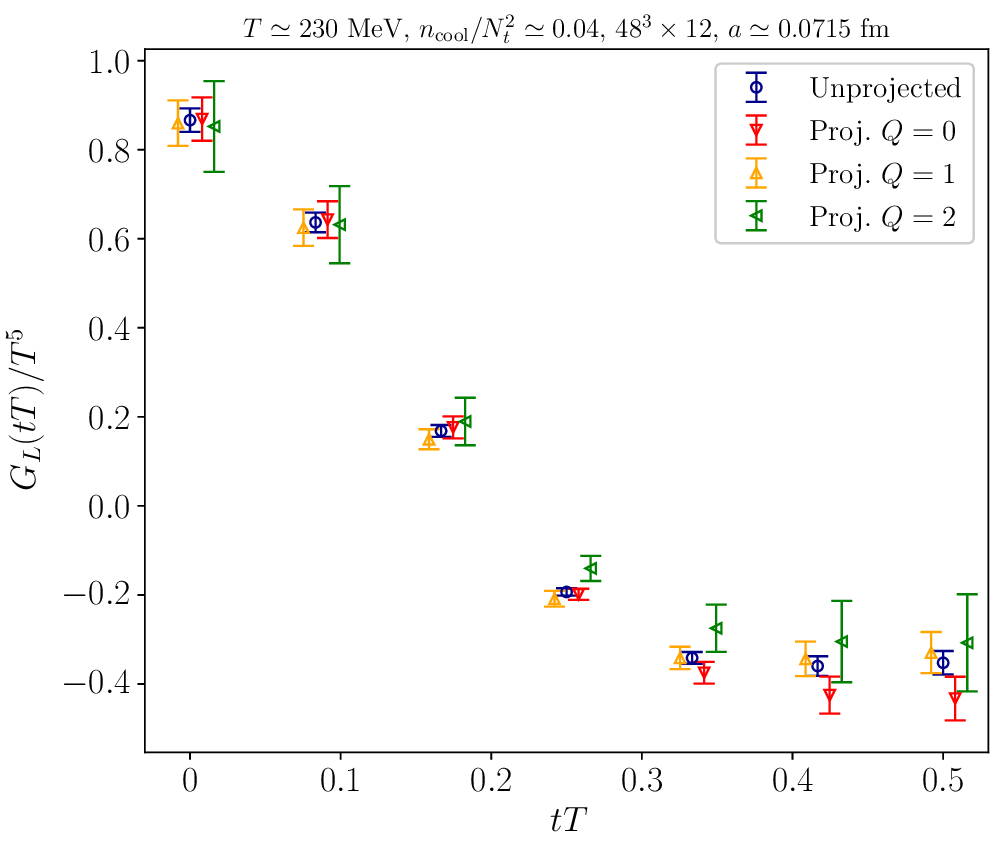}
\includegraphics[scale=0.4]{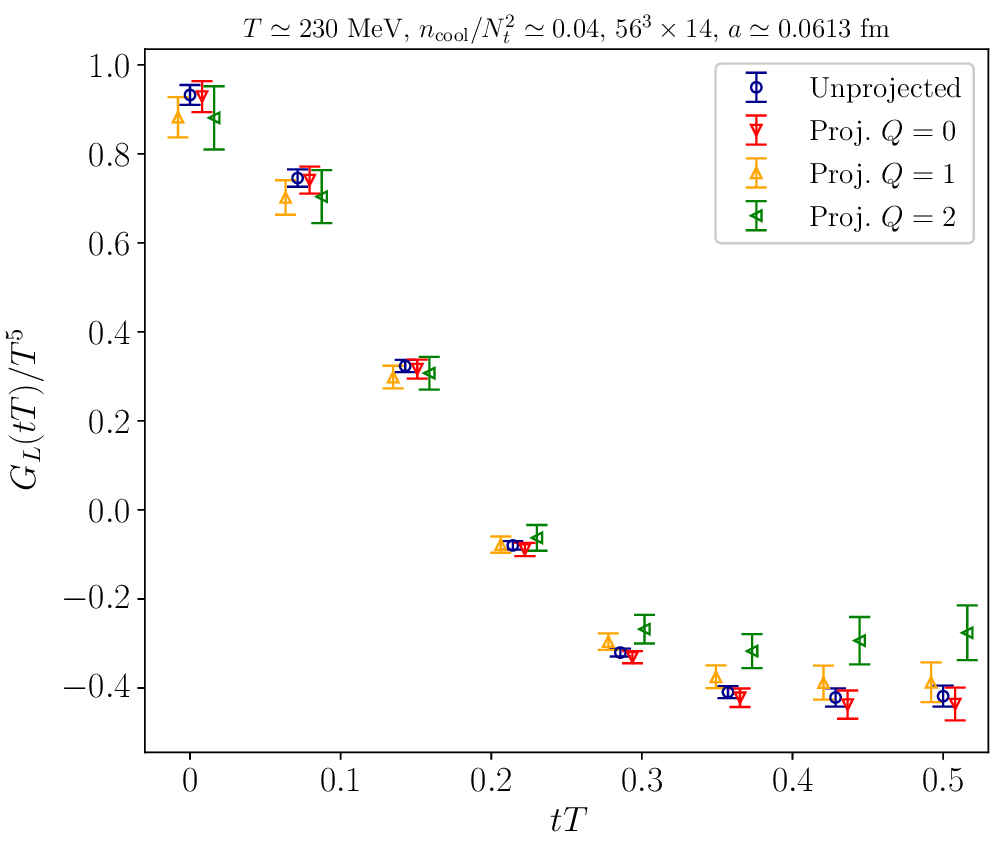}
\includegraphics[scale=0.4]{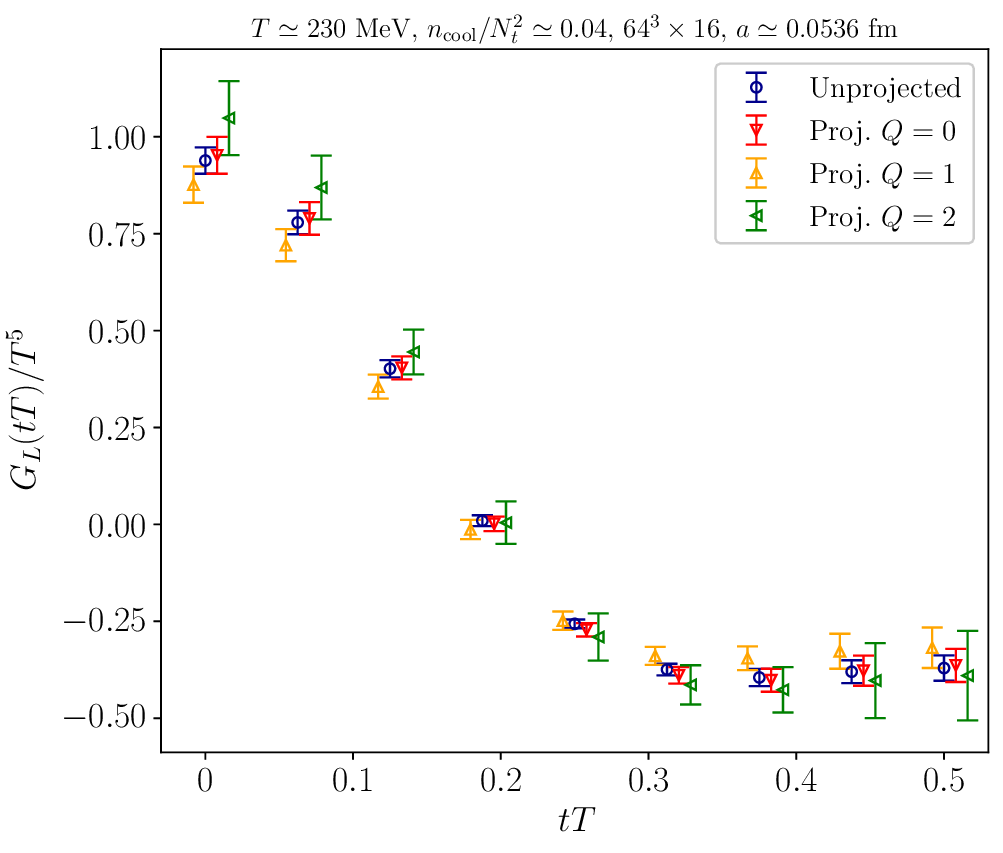}
\caption{Comparison of topological charge density correlator projected in different topological sectors for $T=230$ MeV and $N_t=12$ (top panel), 14 (center panel) and 16 (bottom panel). Plots refer to approximately the same smoothing radius, corresponding to $n_\cool/N_t^2 \simeq 0.04$.}
\label{fig:comp_corr_Qproj}
\end{figure}

\begin{figure}[!t]
\centering
\includegraphics[scale=0.43]{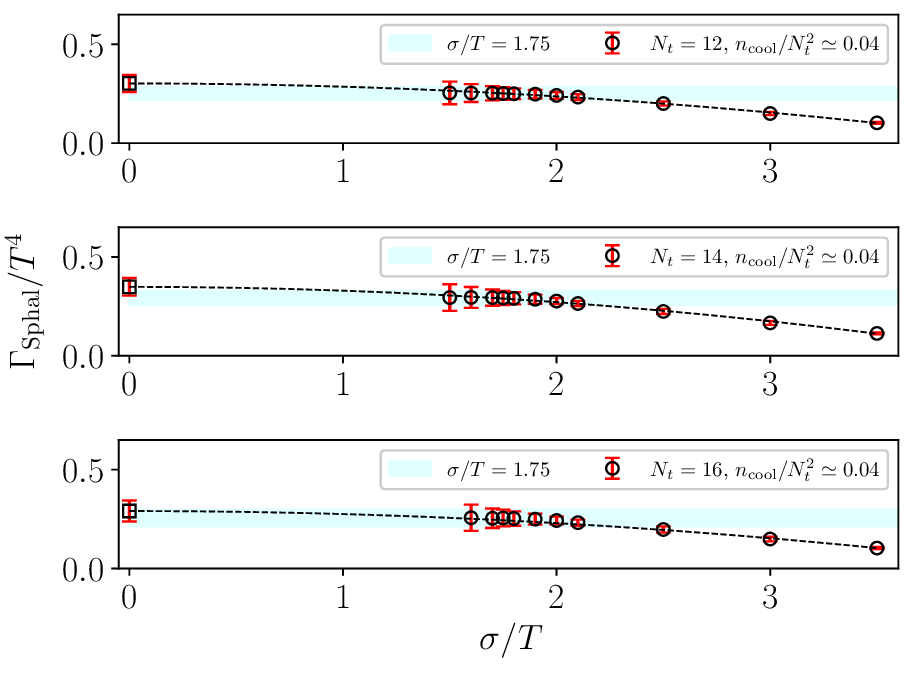}
\includegraphics[scale=0.43]{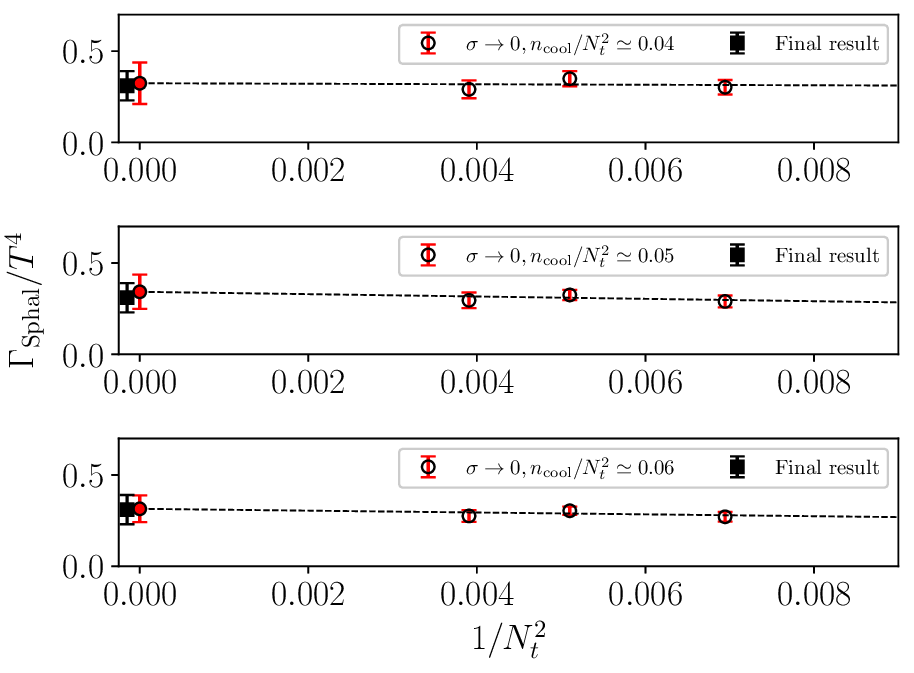}
\caption{Top panel: example of zero-smearing-width extrapolations of the sphaleron rate $\Gamma_\sphal/T^4$ at finite lattice spacing, performed for the three finest lattice spacings explored at $T=230~\mathrm{MeV}$, corresponding to temporal extents $N_t=12,14,16$. Shown examples refer in all cases to the same smoothing radius, corresponding to $n_\cool/N_t^2 \simeq 0.04$, and all extrapolation were performed according to the ansatz in Eq.~\eqref{eq:zerosigma_extr}. Bottom panel: continuum limits of the zero-smearing width extrapolations of the sphaleron rate $\Gamma_\sphal/T^4$ at $T=230~\mathrm{MeV}$ for 3 values of the smoothing radius, corresponding to $n_\cool/N_t^2 \simeq 0.04, 0.05, 0.06$. The obtained continuum extrapolations, reported as filled circle points in $1/N_t=0$, are compared with the final result presented in the main text for this temperature, depicted as filled square point in $1/N_t=0$ as well. In all cases we observe perfect agreement within the errors, very mild lattice artifacts, and no significant dependence on the choice of the smoothing radius.}
\label{eq:check_order_limits}
\end{figure}

Finally, in Figs.~\ref{fig:tcorr_comp_ncool},~\ref{fig:comp_corr_Qproj} and~\ref{eq:check_order_limits} we show further plots referring to $T=230$ MeV.

In Fig.~\ref{fig:tcorr_comp_ncool}, we show the behavior of the correlator for our finest lattice spacing at different values of $n_\cool$, chosen within the range in which we observe a plateau for $\Gamma_\sphal$. We observe that, although the small-$t$ behavior of $G_L(tT)$ is largely affected by the cooling, as expected, this has a small impact on the resulting sphaleron rate. As a matter of fact, we observe the large-$t$ behavior being much less sensitive to the choice of $n_\cool$, which is consistent with the observed plateau in $\Gamma_\sphal$ for smaller values of $n_\cool/N_t^2$.

In Fig.~\ref{fig:comp_corr_Qproj} we compare, for the three finest lattice spacings explored at $T=230$~MeV, the topological charge density correlator obtained with and without projection to a fixed topological sector. The projection in the sector $Q=n$ was performed computing:
\beq
G_n(t) = \frac{\braket{q(t)q(0) \delta(Q-n)}}{\braket{\delta(Q-n)}}.
\eeq
For the cases $n=0,1,2$ we find that the correlators computed projecting in the fixed topological sectors are perfectly compatible within errors among them, thus being also all compatible with the one computed without projection. This observation points out that a bad suppression of non-zero charge topological sectors in the lattice path-integral due to the absence of exact zero-modes of the staggered operator cannot give significant lattice corrections to the correlator, and thus to the sphaleron rate. As a matter of fact, we find that such quantity has indeed very mild corrections to the continuum limit, cf.~Figs.~\ref{fig:T230},~\ref{fig:T300},~\ref{fig:T365},~\ref{fig:T430},~\ref{fig:T570}.

Finally, in Fig.~\ref{eq:check_order_limits} we show that performing the zero-smearing-width extrapolation before the continuum limit, adopting the ansatz in Eq.~\eqref{eq:zerosigma_extr} and keeping the smoothing radius fixed, would yield perfectly agreeing results with those presented in the main text, with no appreciable dependence on the choice of $n_\cool/N_t^2$, and also with no significant lattice artifacts. This is not surprising, as we have shown that, after the continuum limit, the dependence of the obtained results for the sphaleron rate is pretty mild both on $\sigma/T$ and on $n_\cool/N_t^2$, thus pointing out that taking these limits in a different order would yield perfectly compatible determinations of $\Gamma_\sphal$.

\FloatBarrier

\end{document}